# Hardware Acceleration in Portable MRIs: State of the Art and Future Prospects

**Omar Al Habsi[1], Safa Mohammed Sali[1], Anis Meribout[2], Mahmoud Meribout[1] Senior Member IEEE Saif Almazrouei[3], and Mohamed Seghier[3]**

[1]Computer & Information Engineering Department, Khalifa University of Science & Technology, Abu Dhabi, United Arab Emirates (UAE)
[2]Biomedical Engineering and Imaging Institute, Icahn School of Medicine at Mount Sinai (USA)
[3]Biomedical Engineering Department, Khalifa University of Science & Technology, Abu Dhabi, United Arab Emirates (UAE)

Corresponding author: Mahmoud Meriobut (e-mail: mahmoud.meribout@ ku.ac.ae).

**ABSTRACT** There is a growing interest in portable MRI (pMRI) systems for point-of-care imaging, particularly in remote or resource-constrained environments. However, the computational complexity of pMRI, especially in image reconstruction and machine learning (ML) algorithms for enhanced imaging, presents significant challenges. Such challenges can be potentially addressed by harnessing hardware application solutions, though there is little focus in the current pMRI literature on hardware acceleration. This paper bridges that gap by reviewing recent developments in pMRI, emphasizing the role and impact of hardware acceleration to speed up image acquisition and reconstruction. Key technologies such as Graphics Processing Units (GPUs), Field-Programmable Gate Arrays (FPGAs), and Application-Specific Integrated Circuits (ASICs) offer excellent performance in terms of reconstruction speed and power consumption. This review also highlights the promise of AI-powered reconstruction, open low-field pMRI datasets, and innovative edge-based hardware solutions for the future of pMRI technology. Overall, hardware acceleration can enhance image quality, reduce power consumption, and increase portability for next-generation pMRI technology. To accelerate reproducible AI for portable MRI, we propose forming a Low-Field MRI Consortium and an evidence ladder (analytic/phantom validation, retrospective multi-center testing, prospective reader and non-inferiority trials) to provide standardized datasets, benchmarks, and regulator-ready testbeds.

## I. INTRODUCTION

Magnetic Resonance Imaging (MRI) is a non-ionizing imaging modality that uses a static magnetic field and radiofrequency (RF) pulses to excite hydrogen nuclei and measure their returned signals, producing detailed anatomical and functional images across many soft-tissue contrasts. The frequency and phase components of these RF signals are represented in k-space and are converted into 2D/3D MRI images using the Fourier transform [1][2]. MRI is widely used in clinical practice for neurology (stroke, brain tumors, Parkinson's disease), musculoskeletal imaging, cardiology, and oncology because of its excellent soft-tissue contrast and a wide range of acquisition sequences.

Despite its versatility, conventional MRI scanners are bulky, expensive, and require cryogenics, dedicated shielding, and specialized infrastructure, which limits their availability at the point of care [2]. In a typical MRI examination, the preparation phase constitutes a substantial portion of the overall procedure's duration; the patient must be screened for any metallic implants, and any metal accessories must also be removed prior to entering the MRI system room to minimize the risk of harmful incidents caused by the MRI system's intense field [3]. From transporting the patient to initializing the system, the preparation phase can take up to 30 minutes, as shown in [4]. As such, not all medical conditions where time is critical can rely on stationary MRI systems, especially acute conditions where minimizing the diagnosis time is vital for the survival of the patient [5] [6].

Portable MRI (pMRI) systems were developed to address these limitations by bringing imaging to the patient and reducing the need for patient transport. pMRI can enable rapid diagnosis in emergency rooms, ambulances, intensive care units, and resource-limited settings, shortening time-to-diagnosis for time-critical conditions (e.g., acute stroke), increasing accessibility in remote regions, and facilitating longitudinal monitoring where repeated imaging is required. For example, an evaluation was conducted on the Hyperfine Swoop, the only U.S. Food and Drug Administration (FDA) cleared pMRI, to assess its performance compared to conventional MRI scanners (e.g., a MAGNETOM Verio 3T eco and an AVANTO 1.5T) at Yale New Haven Hospital [4]. The comparison involved recording the time required for a complete scan of a patient's head, including the preparation for the intensive care unit (ICU) room, the acquisition of the scan images, and the resetting of the ICU room. Results show that the complete process on Hyperfine Swoop averaged 30:21 minutes compared to 67:36 minutes for conventional high-field MRI scanners.

The value of pMRI extends to resource-limited and remote settings. At Weeneebayko General Hospital in Canada, the deployment of the Hyperfine Swoop improved MRI access for Indigenous communities that otherwise required air transport to distant cities [7], [8]. Similar proposals exist for pMRI use in ambulances [9], the International Space Station [10], and even in regions devastated by war or natural disasters [11]. These examples highlight pMRI's role in expanding imaging access where transportation or infrastructure barriers are prohibitive.

A major limitation of pMRIs is their relatively inferior signal-to-noise (SNR) ratio. This can yield lower image quality or longer scanning times compared to a typical stationary MRI. Recent research strives to improve the performance of pMRI through innovative techniques such as the implementation of artificial learning (AI) models to enhance image quality [12]. Many of these techniques are focused on improving the algorithmic aspect of pMRI, which involves computational models used to improve the diagnostic potential of low-SNR pMRI images.

The integration of MRI with machine learning has opened new diagnostic and therapeutic possibilities. For example, Hussain *et al.* [112] developed a patch-based convolutional neural network that classifies Parkinson's disease from MRI of the substantia nigra using PPMI data, demonstrating the potential of MRI + CNN pipelines for early, automated detection of Parkinsonian changes. Complementing diagnostic applications, adaptive and closed-loop therapies are also being explored: Su *et al.* [113] proposed a fuzzy reinforcement-learning framework for personalizing deep brain stimulation in the presence of communication delay, highlighting how sensing, learning, and control can be combined with imaging and neurostimulation for patient-specific therapy.

As algorithms grow increasingly complex, their computational and hardware demands increase as well. With central processing units (CPUs) approaching their performance limits in recent years, hardware accelerators such as graphics processing units (GPUs), field-programmable gate arrays (FPGAs), and application-specific integrated circuits (ASICs) have emerged. These platforms exploit parallel computing to decompose computationally intensive problems into smaller tasks executed simultaneously across many cores [13], [14]. Since the release of NVIDIA's CUDA library in 2007, the number of publications on GPUs for MRI has surged. [14]. This shows the increasing appeal for hardware acceleration in MRI. There are also developments revolving around the use of FPGAs in MRI image processing [15] [16].

In this paper, we aim to appraise the potential of such hardware acceleration technologies for pMRI. In the past, few reviews on pMRIs were published. They cover various aspects of pMRI technology, including feasibility, challenges, and practicality in the clinical setting. Table I summarizes the selected reviews. Arnold et al. [2] discussed the promise and challenges of pMRI, presenting advantages and disadvantages such as cost, footprint, and power in a comparative table, and highlighting its potential to expand access by complementing high-field MRI. Wald et al. [1] analyzed the cost-effectiveness of pMRI through the history of portable imaging modalities like CT, suggesting its economic viability despite the lack of early clinical examples at the time of publication. Anoardo et al. [17] surveyed innovations to improve low-field MRI, including pre-polarization techniques that temporarily raise the magnetic field to boost SNR up to tenfold while maintaining portability. Kimberly et al. [18] provided a comprehensive overview of brain pMRI, emphasizing hardware, software, and machine learning solutions to improve image quality. Shoghli et al. [6] discussed the role of pMRI in acute neurological conditions such as stroke, noting advantages in emergency use but also risks of false negatives due to low-field limitations. Wilson et al. [19] focused on ultra-low field pMRI in critical care and remote healthcare, offering a broad outlook and recommendations based on a systematic review of over a thousand studies. The review highlighted how pMRI is particularly effective in diagnosing neurological conditions in intensive care units, remote settings, and among specific populations like COVID-19 patients and children.

While past reviews provide an in-depth discussion of various pMRI aspects, a significant gap remains with regard to hardware acceleration, presumably because pMRI is still a new technology, with its first marketable device (The Hyperfine Swoop) released in 2020. To address this gap, we review state-of-the-art pMRI applications with a particular emphasis on back-end technologies, including image reconstruction, hardware acceleration, and related improvements. Although hardware acceleration for pMRI is still in its infancy, it is expected to play a major role in future systems. The main contributions of this paper are:

1) A survey of the state-of-the-art hardware acceleration in pMRIs.

**TABLE I: REVIEW ARTICLES ON PMRIs**

| Reference | Title | Year | Highlights |
|---|---|---|---|
| [2] | Low-field MRI: Clinical promise and challenges | 2022 | Focused on specific clinical applications |
| [1] | Low-Cost and pMRI | 2019 | Cost-oriented |
| [17] | New challenges and opportunities for low-field MRI | 2022 | Front-end focused |
| [18] | Brain imaging with portable low-field MRI | 2023 | General Overview |
| [6] | Current role of pMRI in diagnosis of acute neurological conditions | 2023 | Focused on acute neurological conditions |
| [19] | Applications, limitations and advancements of ultra-low- field magnetic resonance imaging: A scoping review | 2024 | Systematic literature review of 25 ULF MRI studies |
| Ours | Hardware Acceleration in Portable MRIs: State of the Art and Future Prospects | 2025 | Focuses on hardware accelerators in pMRI; discusses its real-time deployment and challenges |

2) A review of the state-of-the-art image processing algorithms in pMRIs.
3) A discussion of future developments for the back-end section on MRIs.

Figure 1 illustrates the flow diagram of literature search strategy. Hundreds of papers were reviewed in Scopus and Google Scholar databases using inclusive keywords: “pMRI”, “real-time pMRI”, and “Hardware accelerator for pMRI” for the duration of 2020-2025. The keyword “MRI” was excluded since the focus was on pMRI only. Literature that uses hardware accelerators like GPU, FPGA and ASIC for functions other than accelerators were filtered. Hardware accelerators for conventional MRI have contributed to achieving functional MRI, which allows the display of organs in motion. This is not possible with p-MRI where the image formation is dominated by the slow image reconstruction. Hence, the role of the hardware accelerator is to reduce the image reconstruction time to few ms, without achieving real-time performance.

Section II provides background information on pMRI technologies. Section III introduces the concepts and algorithms that underpin image reconstruction. Section IV discusses different types of hardware accelerators and their potential applications for pMRI. Section V presents a review of selected pMRI studies. Finally, Section VI offers a discussion of the key challenges and opportunities in this field.

## II. BACKGROUND

### A. THE ANATOMY OF A PORTABLE MRI

Elementary particles possess a spin that determines their magnetic properties. When placed in a magnetic field, they process (or rotate) around an axis in the direction of, or opposite to, the direction of the magnetic field [21] [22]. Figure 2 presents a pictorial representation of protons aligning their spin axes when subjected to a static magnetic field.

The frequency at which the nuclei process is called the Larmor Frequency ($f_0$) and is dependent on two factors: the static magnetic field ($B_0$), and the gyromagnetic ratio of the particle ($\gamma$), a constant determined by the type of nucleus.

$$f_0 = \gamma B_0 / 2\pi \quad (1)$$

Because of the use of ultra-low magnetic fields ($B_0$ less than 0.1T), pMRI operates at very low Larmor frequencies. Like conventional MRI, by contrasting the differences in relaxation times of the processed hydrogen atoms present in water, pMRI can construct a detailed anatomical image [21]. The following subsections provide an outline of the subsystems in a typical pMRI system’s design.

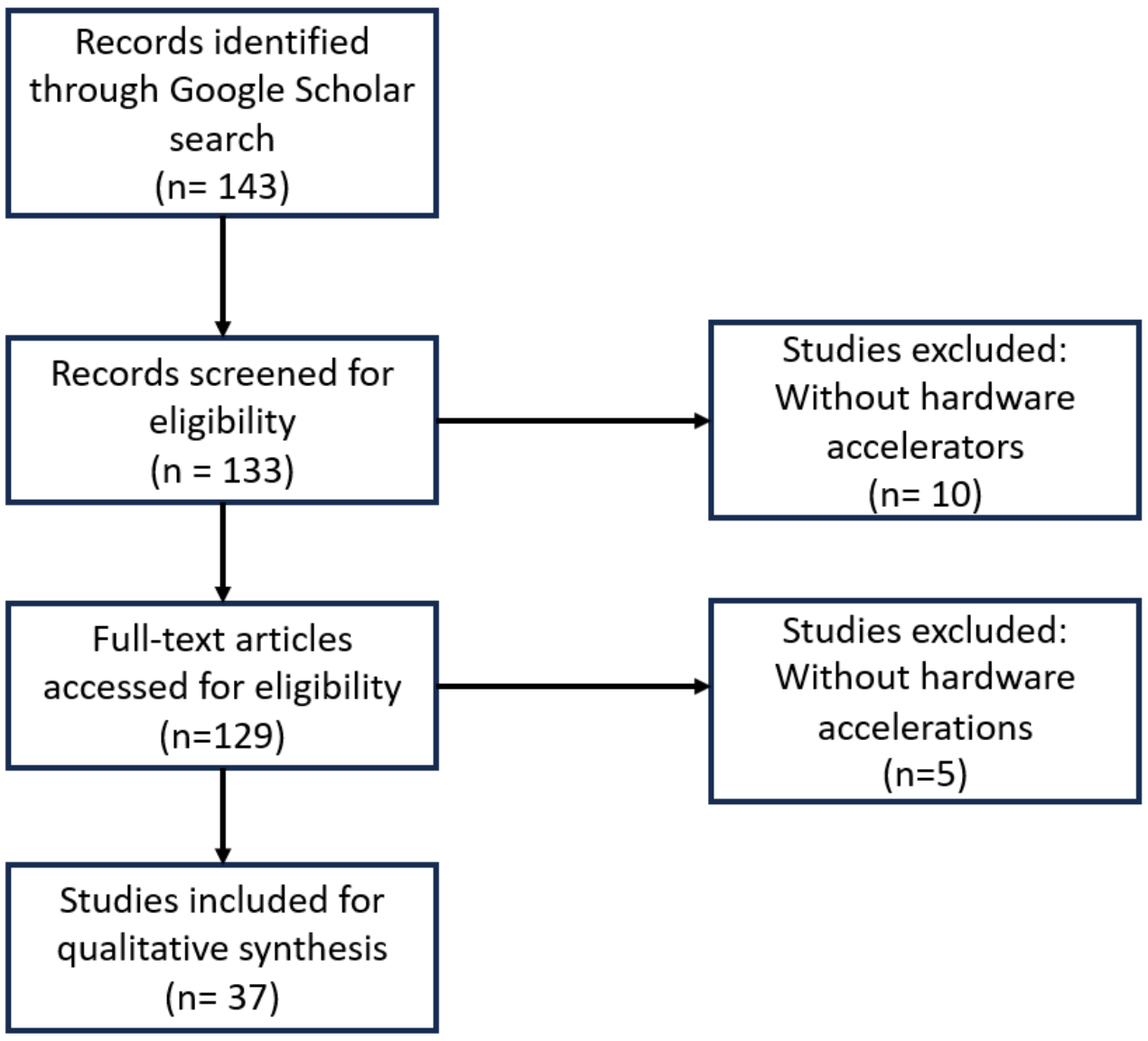

FIGURE 1. Flow diagram of literature search strategy.

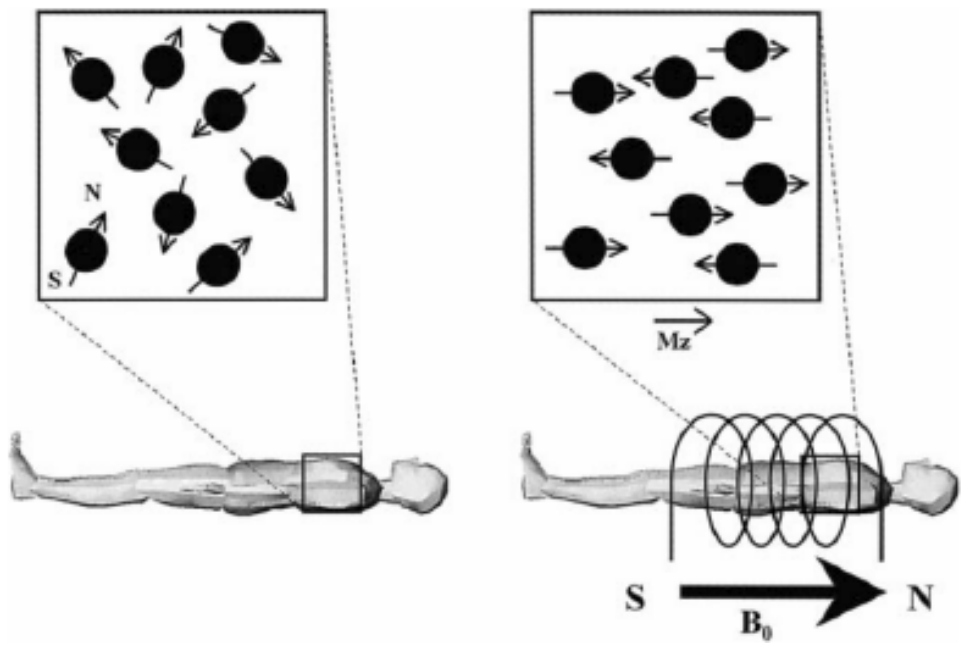

**FIGURE 2. Protons precessing along the Mz axis when subjected to a static magnetic field, $B_0$, similar to a compass. [21]**

### 1) MAGNET

The magnet is responsible for aligning the hydrogen atoms' spins and accounts for much of the system's weight. It generates the static magnetic field, $B_0$, which determines system performance. Key design considerations include field strength, weight, homogeneity, and cost. Field strength directly influences image signal-to-noise ratio (SNR) and resolution, while weight affects portability. Homogeneity is critical to minimizing image artifacts, and the cost varies significantly depending on the design and cooling requirements. MRI magnets are generally classified into three types: permanent magnets, resistive coils, and superconducting coils. Permanent magnets use ferromagnetic materials to generate constant magnetic fields without power or cooling, making them ideal for pMRI systems [18].

While energy- and cost-efficient, their field strength is limited (50–200 mT) [6]. Common materials include Neodymium-Iron-Boron (NdFeB) and Samarium-Cobalt (SmCo). NdFeB provides stronger fields and lower costs but is prone to temperature instability and corrosion, requiring protective coatings. SmCo offers greater thermal stability (lower field degradation: 0.015%/°C vs. NdFeB's 0.125%/°C) and durability, but at a higher cost and weight [23] [24] [25] [26] [27]. Resistive coils require power and cooling, operating as air-cooled systems at low field strengths (up to 60 mT). They are primarily used for extremity imaging in low-field MRI and often employ copper coils in solenoid or Helmholtz configurations to enhance homogeneity [17] [28] [29]. Superconducting coils, common in high-field MRI, achieve the highest $B_0$ values (1000 + mT). However, cryogenic cooling requirements hinder portability. Research on low-field superconducting systems remains limited, with emerging studies exploring high-temperature superconducting coils to address this challenge [30][31].

### 2) SHIMMING

Shimming refers to the process of improving the homogeneity of $B_0$ in an MRI system. Field inhomogeneities arise due to imperfections in the main magnet and variations introduced by the subject's anatomy, which can lead to artifacts and degraded image quality. Shimming is achieved by superimposing additional magnetic fields over $B_0$ to cancel these variations [32] Depending on how these corrections are implemented, shimming is classified into passive, active, and dynamic approaches.

Passive shimming involves the strategic placement of permanent magnets or ferromagnetic materials in the inner bore of the magnet to provide static corrections to $B_0$ [33]. In contrast, active shimming utilizes electromagnetic coils that dynamically adjust the field corrections by applying a varying current to coils located in the inner bore of the magnet [32]. Active shimming entails recurrently mapping the $B_0$ field, identifying the target region, measuring the local irregularities, and calculating the currents of each coil that would minimize the error [34]. Since this is an inverse problem, it is computationally demanding and can introduce a delay in scan times. For instance, using a tool called Shimming Toolbox, running a dynamic shimming procedure required around 5 minutes [34]. In this context, hardware acceleration can play a pivotal role in making these active shimming processes faster and more adaptable, pushing the boundaries of what can be achieved with shimming technologies in pMRI.

### 3) GRADIENT

Gradients are essential components of an MRI system, responsible for spatial encoding of the magnetic resonance signal [35]. They are generated by gradient coils that superimpose a linearly varying magnetic field onto the static magnetic field ($B_0$), and they have a direct impact on the spatial resolution, accuracy, and speed of imaging. Ensuring accurate gradients is vital for high-quality imaging because imperfections in gradients can introduce artifacts like geometric distortions [36]. To ensure high gradient performance, both hardware and software solutions are employed. Hardware improvements focus on the design of gradient coils with optimized geometry to improve linearity and reduce artifacts [37]. Software-based techniques complement hardware enhancements by calibrating gradient fields through pre-scanning routines and real-time corrections. Algorithms are used to model the imperfections in gradient behavior and apply adjustments to ensure accurate spatial encoding during imaging [38].

Hardware acceleration can enhance the performance of software-based gradient corrections. Technologies such as GPUs and FPGAs enable rapid computation of eddy current corrections and gradient nonlinearity calibrations, reducing the processing time for these adjustments. In dynamic imaging applications, hardware-accelerated processing allows real-time gradient corrections, ensuring that high-speed imaging techniques remain accurate and artifact-free. These accelerators also facilitate the integration of machine learning algorithms, which can predict and correct gradient-induced distortions more efficiently, further improving image quality [39]. In sum, advancements in gradient hardware design, software-based calibration techniques, and hardware-accelerated processing ensure precise gradient performance, reducing artifacts and enhancing the spatial and temporal resolution of pMRI images.

#### 4) RADIO FREQUENCY COILS

RF (Radiofrequency) coils generate RF pulses to excite hydrogen protons and detect the emitted MR signals, directly influencing image quality, including signal-to-noise ratio (SNR), spatial resolution, and uniformity. RF coils are classified into transmit coils, receive coils, and transmit-receive (T/R) coils. Transmit coils, like body coils, generate RF pulses, while receive coils, such as phased-array surface coils, capture MR signals with high sensitivity [40] [41] [42] [43]. Transmit-receive coils combine both functions, typically in smaller applications. Coil design is tailored for specific regions like the head, spine, or extremities to optimize coverage and SNR [40] [42].

Key performance parameters include SNR, homogeneity, and specific absorption rate (SAR). SNR measures signal clarity, homogeneity ensures uniform RF field (B1) distribution, and SAR quantifies energy absorption, which must be managed for patient safety [42] [44] [45]. Imperfections like non-uniform B1 fields, coil coupling, and high SAR levels can introduce artifacts, signal dropouts, or intensity variations [41] [44]. Advances in RF coil technology, such as multi-channel phased-array coils, improve SNR and enable parallel imaging through independent receive elements [40]. Modern RF coils often include multiple receiving coils that enable parallel imaging techniques, which, through hardware acceleration, can effectively reduce the acquisition time and computational load of the imaging process [42] [44].

### B. MEDICAL STANDARDS

In the U.S., the FDA oversees the approval process of new medical devices using a risk-based classification system to ensure their safety and effectiveness. Currently, there are 3 known classes of medical devices: Class I (low risk), Class II (moderate risk), & Class III (high risk). While all classes are required to follow general control provisions such as device registration & misbranding policies, Class II & Class III are subjected to special controls due to their higher risk profiles [47]. Since 1988, MRIs have been reclassified from Class III into Class II [46]. Class II devices are required to submit a 510(k) premarket notification to acquire clearance for commercial distribution. In the notification, the manufacturer must demonstrate that the new product (subject device) is substantially equivalent to an approved product (predicate device). Proving substantial equivalence involves comparing the intended use, technological characteristics, and performance data of both devices. MRI systems typically share the same intended use and technological characteristics as conventional MRI systems, making them comparable under these criteria.

Although optional, the FDA encourages the declaration of adherence to one or more of its recognized standards to improve the likelihood of approval. These performance standards include the International Electrotechnical Commission Standard (IEC) 60601-2-33 [48], which ensures the basic safety and essential performance of MRI equipment, and the National Electrical Manufacturers Association (NEMA) standards, which require specific performance metrics including SNR [49] and acoustic noise [50]. Given the diverse settings in which pMRI systems are used, and compared to conventional MRI systems, certain standards can be more challenging to meet. General safety standards, such as IEC 60601, require demonstrating product safety across various clinical environments. Furthermore, the ultra-low field $B_0$ in pMRI systems presents challenges in complying with performance standards like NEMA MS 1 and NEMA MS 4, which emphasize metrics such as SNR and image quality.

### C. COMPARATIVE METRICES

A structured evaluation of pMRI systems calls for a set of quantitative measures that can consistently capture image quality, speed, efficiency, and portability across diverse designs. This ensures that T1W, T2W, FLAIR, and other acquisitions can be evaluated on a consistent scale, enabling meaningful cross-system analysis rather than a purely descriptive listing. The following are the comparative matrices used in this paper:

#### 1) PEAK SIGNAL-TO-NOISE RATIO (PSNR)

In portable MRI, where undersampling and low-field noise can introduce significant errors, PSNR provides a straightforward way to quantify how much the reconstruction deviates from a reference. It is derived from the mean squared error (MSE) between a reconstructed image and a reference image.

$$PSNR = 10log_{10}\left(\frac{MAX_I{}^2}{MSE}\right) \quad (2)$$

Where $MAX_I$ is the maximum possible pixel intensity in the image (for example, 255 for an 8-bit image or 1.0 for normalized MRI data), and

$$MSE = \frac{1}{N}\sum_{i=1}^{N}(x_i - y_i)^2 \quad (3)$$

is the average squared difference between the reference image $x$ and the reconstructed image y, over all N pixels.

To measure PSNR in MRI, a fully sampled or high-field MRI image is typically used as the reference. The portable or accelerated reconstruction is scaled to the same intensity range as the reference. Then MSE is computed pixel by pixel, and PSNR is calculated using the formula above. Higher PSNR values indicate closer agreement with the reference image. PSNR values above 30 dB are usually regarded as good, with values above 40 dB corresponding to near-indistinguishable reconstructions.

#### 2) STRUCTURAL SIMILARITY INDEX (SSIM)

The structural similarity index is a widely adopted perceptual metric designed to evaluate the fidelity of a reconstructed image relative to a reference. SSIM models the way humans perceive image degradation by incorporating three complementary components: luminance (brightness), contrast, and structure. For two images $x$ and $y$, SSIM is computed over local patches as:

$$SSIM(x, y) = \frac{(2\mu_x\mu_y + C_1)(2\sigma_{xy} + C_2)}{(\mu_x{}^2 + \mu_y{}^2 + C_1)(\sigma_x{}^2 + \sigma_y{}^2 + C_2)} \quad (4)$$

where $\mu_x$ and $\mu_y$ are the mean intensities of the patches, $\sigma_x{}^2$and $\sigma_y{}^2$ their variances, $\sigma_{xy}$ their covariance, and $C_1, C_2$ re small constants that stabilize division. The final SSIM score is the mean of all patch-level scores. Values range from 0 (no similarity) to 1 (perfect structural similarity).

To measure SSIM in MRI, a high-quality image, often obtained with fully sampled k-space data or a validated high-field reference scan, serves as the ground truth. The reconstructed image from the portable or accelerated system is compared patch-by-patch with the reference. SSIM is then averaged over the entire volume or over clinically relevant slices. An SSIM value above 0.9 is typically considered excellent, while values below 0.8 may indicate visible structural loss. SSIM penalizes distortions that alter spatial patterns and anatomical structures, making it particularly relevant for diagnostic imaging, where preserving subtle features (e.g., lesions or cortical folds) is critical. SSIM is also less sensitive to uniform brightness shifts, which are common in MRI due to coil bias fields, meaning it aligns better with clinical perception of image quality.

#### 3) DIAMETER OF SPHERICAL VOLUME (DSV)

It is measured from the magnet bore size and defines the largest volume over which the magnetic field is sufficiently homogeneous for imaging.

#### 4) MRI ACQUISITION SEQUENCES

MRI protocols differ based on how tissue contrast is emphasized, which directly influences scan time, resolution, and diagnostic utility. In portable and low-field MRI systems, commonly used sequences include:

a) T1-weighted (T1W): Produces images where fat appears bright and cerebrospinal fluid (CSF) appears dark, making it valuable for detailed anatomical visualization and structural assessment.
b) T2-weighted (T2W): Highlights water-rich tissues such as edema, inflammation, and tumors, with CSF appearing bright, making it effective for pathology detection.
c) FLAIR (Fluid-Attenuated Inversion Recovery): A T2-derived sequence where the CSF signal is suppressed, enhancing visibility of lesions near ventricles, such as multiple sclerosis plaques or strokes.
d) DWI (Diffusion-Weighted Imaging): Sensitive to water molecule diffusion, with restricted diffusion (e.g., in acute ischemic stroke) appearing bright, making it indispensable for early stroke diagnosis.
e) PDW (Proton Density-Weighted): Reflects the number of hydrogen nuclei in tissues, providing intermediate contrast useful for distinguishing subtle anatomical differences.

Their inclusion explains the variability in scan times and voxel resolutions across systems, and forms the basis for subsequent quantitative comparison.

## III. RECONSTRUCTION ALGORITHMS

### A. THE FOURIER TRANSFORM OF THE k-SPACE

In MRI, the inverse Fast Fourier Transform (iFFT) can reconstruct spatial-domain images from acquired k-space data,which is a spatial frequency-domain representation of the scanned region. The iFFT is fundamental in MRI reconstruction, especially under Cartesian sampling conditions, due to its simplicity, accuracy, and computational efficiency. For atwo-dimensional Cartesian grid, the reconstructed image is given by:

$$I(x, y) = \sum_{k_x=0}^{N-1} \sum_{k_y=0}^{M-1} K(k_x, k_y) e^{j2\pi\left(\frac{k_x x}{N} + \frac{k_y y}{M}\right)} \quad (5)$$

where $K(k_x, k_y)$ is sampled k-space data and $I$(x, y) represents the image intensity at position ($x$, $y$). This transformation efficiently converts frequency-domain data into spatial-domain images. The FFT algorithm reduces the computational complexity of the DFT from $O(N^2)$ to $O(N\log N)$. The uniform sampling under Cartesian sampling allows the iFFT to be applied without additional preprocessing, simplifying the reconstruction process. However, its applicability is limited when k-space data is acquired non-uniformly, as in advanced imaging techniques such as radial or spiral sampling. These require the use of additional techniques such as gridding, where k-space data is interpolated onto a uniform grid prior to iFFT, or the use of a more computationally advanced variant of FFT called the Non-uniform Fast Fourier Transform (NuFFT). Such sampling variants might add computational costs during image reconstruction.

### B. COMPRESSED SENSING

Compressed sensing (CS) is an advanced signal processing technique that allows for the reconstruction of high-resolution images from fewer data samples than typically required [69]. This is achieved by exploitingthe sparsity of the MRI data, enabling the reconstruction of images from undersampled datasets [70]. By integrating CS with advanced reconstruction techniques, MRI systems can achieve faster imaging without compromising image quality, which is particularly beneficial in dynamic imaging scenarios.While computationally intensive, CS is effective in reducing scan times; as illustrated in Figure 2, a prior study showed thatCS can recover most relevant features of an image even under 10-fold undersampling [70]. Figure 3 depicts the mentioned result.

### C. PARTIAL FOURIER IMAGING

Partial Fourier sampling leverages the conjugate symmetry property of k-space data, allowing for the reconstruction of complete images from partially acquired data [71]. In MRI, k-space data exhibit Hermitian symmetry, meaning that one half of the k-space contains sufficient information to infer the other half. By acquiring slightly more than half of the k-space data, typically about 60% and employing reconstruction algorithms to estimate the missing data, this technique effectively reduces acquisition time. However, due to potential phase errors arising from factors like magnetic field inhomogeneities or motion, additional data are often sampled to ensure accurate reconstruction [72].

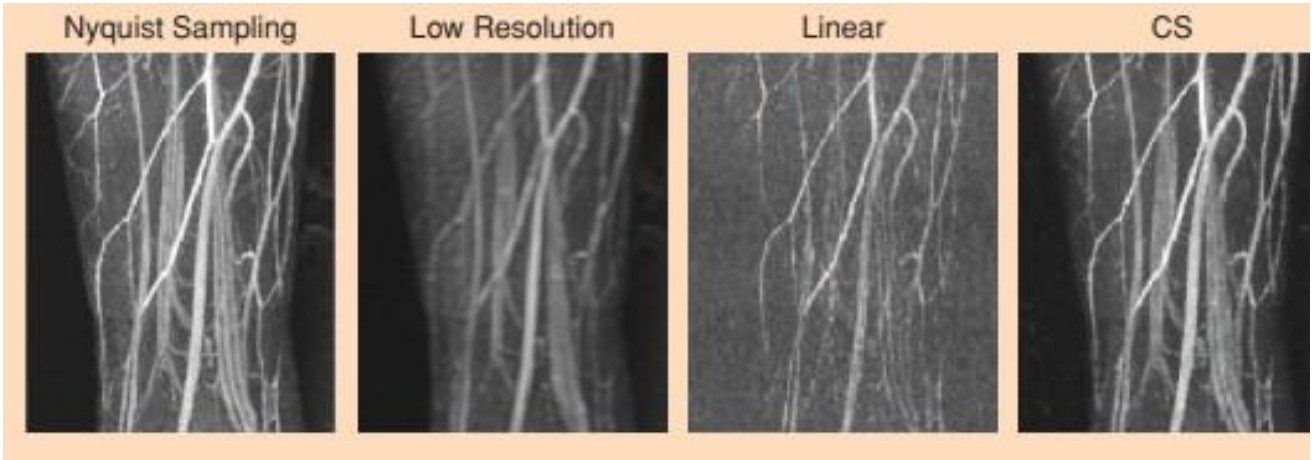

**FIGURE 3.** **3-D Contrast-enhanced angiography. Even with 10-fold undersampling, CS can recover most blood vessel information revealed by Nyquist sampling; there is significant artifact reduction compared to linear reconstruction, and a significant resolution improvement comparedto a low-resolution centric k-space acquisition. Images used from the following cited source [70].**

### *D. PARALLEL IMAGING*

As was mentioned in Section II.A.4, parallel imaging can accelerate data acquisition using multiple receiver RF coils. By undersampling k-space data, parallel imaging effectively reduces the scan time. Specialized algorithms process the undersampled data to reconstruct the full image while maintaining image quality. However, the resultant improvement in scan time comes with a trade-off in reduced signal-to-noise ratio (SNR) [51]. The reduction in SNR due to parallel imaging is quantified by the equation:

$$SNR_r^{accelerated} = \frac{SNR_r^{full}}{g_r\sqrt{R}} \quad (6)$$

where $SNR_r^{\text{accelerated}}$ is the SNR after parallel imaging reconstruction, $SNR_r^{\text{full}}$ is the SNR without acceleration, $R$ is the acceleration factor, and $g_r$ is the geometry factor, representing noise amplification due to coil geometry and reconstruction process [52]. This relationship indicates that as the acceleration factor $R$ increases, the SNR decreases proportionally, compounded by the geometry factor $g_r$. Parallel imaging reconstruction algorithms have an inherent capability to be parallelized computationally [53]. The reconstruction process can be further accelerated by exploiting the independence of the data operations and breaking it down across multiple processors. The aforementioned algorithms are all parallelizable. For instance, k-t SPARSE combines parallel imaging and CS to reconstruct images from the under-sampled k-space data [110]. The algorithm uses a set of training data acquired from the same subject to estimate the missing k-space data in the under-sampled image, and it incorporates temporal information to accurately capture the dynamic images of the organs [111].

There are also two widely used parallel imaging reconstruction algorithms: Sensitivity Encoding (SENSE) and Generalized Autocalibrating Partially Parallel Acquisitions (GRAPPA).

#### 1) SENSITIVITY ENCODING (SENSE)

SENSE is a parallel-imaging method that reconstructs undersampled data in the image domain by exploiting spatially varying coil sensitivity profiles. The forward model is succinctly written as

$$m = S \times I + n \quad (7)$$

where $m$ represents the measured signals from all coils in the image domain, $S$ is the coil sensitivity matrix, $I$ is the vector of the true image to be reconstructed, and $n$ accounts for noise in the system. To reconstruct image I, SENSE solves the inverse problem:

$$I = (S^H S)^{-1} S^H m \quad (8)$$

where $S^H$ is the conjugate transpose of the sensitivity matrix.

SENSE's practical appeal for accelerator design stems from this arithmetic regularity and locality: each pixel (or small neighbourhood) requires only a few matrix multiplications and a low-dimensional inversion, yielding excellent data-parallelism that maps naturally to FPGAs and GPUs. However, SENSE depends on accurate sensitivity estimation; calibration errors (motion, variable loading, or simplified coil arrays common in portable/low-field systems) propagate into structured artifacts and limit robustness. Preprocessing for sensitivity estimation, whether performed online or offline, adds to end-to-end latency and must be included in real-time performance analysis.

System-wise, SENSE is best suited for Cartesian sampling and scenarios demanding rapid, low-latency reconstruction with reliable calibration. Hardware studies confirm SENSE's accelerator affinity: optimized pipelines on Xilinx Zynq UltraScale+ platforms achieved up to 298× speed-up over CPU baselines for 8-coil datasets, reconstructing images in 2.96 ms versus 190 ms (CPU) [15].

Scalability remains the main bottleneck. Larger coil arrays and higher acceleration factors worsen conditioning, demanding additional numerical stabilization and hardware resources. These constraints are especially relevant in portable MRI, where space and calibration fidelity are limited.

#### 2) GENERALIZED AUTOCALIBRATING PARTIALLY PARALLEL ACQUISITIONS (GRAPPA)

GRAPPA reconstructs missing k-space samples by linearly combining neighboring acquired k-space points across coils, using a fully sampled central region (the auto-calibration signal, ACS) to estimate convolutional weights. The calibration and synthesis are compactly written as

$$\boldsymbol{K}_{ACS,\,pred} = \boldsymbol{K}_{ACS,\,adj}\boldsymbol{W} \quad (9)$$

where $\boldsymbol{K}_{ACS,\,pred}$ represents the k-space points to be predicted within the ACS region, $\boldsymbol{K}_{ACS,\,adj}$ contains adjacent k-space samples and $\boldsymbol{W}$ is the weight matrix obtained by a least-squares calibration. For example,

$$\boldsymbol{W} = \left(\boldsymbol{K}_{ACS,adj}{}^{H}\boldsymbol{K}_{ACS,adj}\right)^{-1}\left(\boldsymbol{K}_{ACS,adj}{}^{H}\boldsymbol{K}_{ACS,pred}\right) \quad (10)$$

where $\boldsymbol{K}_{ACS,adj}{}^{H}$ is the conjugate transpose of $\boldsymbol{K}_{ACS,adj}$. Once calibrated, $\boldsymbol{W}$ is convolved across the undersampled k-space to synthesize missing lines and produce a full k-space for inverse FFT.

Unlike SENSE, GRAPPA does not require explicit coil-sensitivity maps, increasing robustness to calibration errors and making it attractive in portable systems with simplified or variable coil geometries. This resilience comes at the cost of heavier arithmetic: calibration involves dense matrix operations and the synthesis step performs repeated convolutions over k-space. From an accelerator viewpoint, these operations are both a strength and a challenge — the convolutional and matrix-multiply kernels map well to parallel hardware (systolic arrays, SIMD engines, or FPGA DSP blocks), but they impose high demands on memory bandwidth and on-chip buffering to sustain throughput.

In practice, GRAPPA implementations have been successfully accelerated. FPGA prototypes on Xilinx Zynq UltraScale+ platforms (ZCU102) report speed up to 121× compared to CPU and 9× faster than the GPU-based Gadgetron [55] framework [16]. The FPGA reconstructed a 160×120 frame in 48–53 ms, while the CPU and GPU reconstructed the same frame in 6,416 ms and 437 ms, respectively.Thus, GRAPPA offers a practical trade-off for portable MRI: greater robustness to coil-map errors at the expense of increased compute and memory requirements, which must be provisioned or mitigated (e.g., via kernel fusion, fixed-point arithmetic, or hierarchical buffering) in edge accelerator designs.

### 3) ARTIFICIAL INTELLIGENCE-BASED METHODS

Deep learning (DL) has transformed MRI reconstruction by learning data-driven mappings from undersampled k-space to high-quality images. Unlike SENSE or GRAPPA, which rely on explicit forward models, DL methods adaptively capture complex priors, enabling higher acceleration factors with reduced artifacts and noise [56]. Their impact extends beyond reconstruction to pre-processing (e.g., learned gridding for non-Cartesian sampling [58]) and post-processing (denoising and super-resolution [59], [60]), making them increasingly central to modern MRI pipelines.

The practical relevance for portable MRI lies in two dimensions: robustness and hardware efficiency. AI-driven reconstructions are often more tolerant of noisy, low-field data than traditional model-based methods, preserving diagnostic structures even under aggressive undersampling. At the same time, inference workloads (dominated by convolutions) map naturally to GPUs, FPGAs, and ASICs, aligning with the parallel architectures already deployed in edge accelerators. This makes real-time AI-enhanced reconstruction feasible in compact systems, as demonstrated by CNN-based FPGA implementations achieving millisecond-scale throughput [62], [63].

#### a) Convolutional Neural Networks (CNNs)

CNNs have become the dominant architecture for MR reconstruction because their convolutional kernels map well to the local structure of k-space and image-domain features. Their practical strengths for portable MRI include:

- Hardware affinity: Convolutions parallelize well on FPGA DSP blocks and embedded GPUs; depthwise/separable convolutions and MobileNet-style blocks reduce arithmetic intensity and memory. Kernel fusion, streaming dataflows, and reuse of intermediate buffers are critical FPGA/SoC optimizations to meet real-time targets. For example, a MobileNetV2 implementation on an Intel Arria 10 SoC FPGA demonstrated 266.6 frames-per-second (≈3.75 ms per image), a ≈20× speedup over a CPU baseline, by exploiting depthwise and pointwise convolution parallelism and streaming execution [62]. In parallel, reconstruction studies have shown that CNN architectures can achieve state-of-the-art image quality. For instance, at 75% undersampling, PSNR values up to 34.33 dB and SSIM ≈ 0.928 were reported for 256×256 MR images processed on an Intel Core i7-7700 CPU and GeForce Titan XP GPU [63].
- Compression readiness: Pruning, quantization, and separable convolutions reduce memory and energy costs significantly with only minor accuracy loss, making CNNs suitable for edge deployment where power and cooling are constrained.
- Integration with physics: Unrolled or plug-and-play architectures that embed data consistency/FFT steps achieve higher accuracy per parameter and are more robust to acquisition variations than purely black-box CNNs; they also reduce the total number of learned parameters, easing hardware requirements.

However, limitations remain. CNNs often require large, paired training datasets that match the target hardware and coil setup; data rarely available for low-field systems. Domain shift (different field strengths, patient populations, coil layouts) can lead to degraded reconstructions if not addressed through transfer learning or self-supervised strategies. Moreover, while CNNs can deliver impressive PSNR/SSIM values under simulated undersampling, these do not always translate to consistent diagnostic performance in real-world portable scans. Thus, evaluation must go beyond pixel metrics to task-specific outcomes (lesion detection, volume measurement).

#### b) Generative Adversarial Networks (GANs)

GANs extend MRI reconstruction beyond pixel-wise fidelity by explicitly modeling perceptual realism. A generator synthesizes candidate reconstructions from undersampled data, while a discriminator enforces anatomical plausibility by penalizing unrealistic textures and structures. This adversarial training encourages the network to capture fine-grained features that conventional CNN or parallel-imaging reconstructions tend to oversmooth. The outcome is not only improved quantitative scores (PSNR, SSIM) but also enhanced perceptual quality, which is particularly critical in low-field and portable MRI where SNR is intrinsically limited. For example, recent studies report SSIM gains up to 0.85–0.90 and PSNR improvements of 2–7 dB over conventional methods when GANs are applied to low-field reconstructions [67].

From a hardware perspective, GANs are more challenging to deploy than CNNs due to the presence of two coupled networks and the heavy reliance on transposed convolutions in the generator. These layers are bandwidth-intensive and inefficient on conventional convolution accelerators. To address this, FPGA-specific designs such as FlexiGAN [68] restructure the generator to align with FPGA-friendly primitives, demonstrating a 2.2× performance improvement on a Xilinx XCVU13P device compared to generic convolution accelerators, and a 2.6× better performance-per-watt than an NVIDIA Titan X GPU. Although FlexiGAN was evaluated outside MRI, the design principles directly translate to pMRI reconstruction: rebalancing compute between generator and discriminator, streaming intermediate activations, and exploiting FPGA parallelism to meet real-time constraints.

GAN-based reconstruction provides a pathway to clinically usable image quality in noisy, low-field MRI by enhancing perceptual fidelity, but its hardware realization demands careful architectural tailoring. Emerging FPGA frameworks show that with generator-specific optimizations, adversarial reconstruction can be accelerated efficiently enough for portable deployment, which is a key enabler for field-ready pMRI systems.

## IV. HARDWARE ACCELERATORS

As medical technologies have become more sophisticated, the computational demand to operate these technologies has increased at levels sometimes beyond typical Central Processing Units' (CPUs) processing capabilities. Consequently, hardware acceleration has been receiving more interest as a means to overcome this computational challenge [73]. In broad terms, hardware acceleration is a technique to offload complex computational tasks from the CPU to a specialized processing unit called hardware accelerators; this allows the CPU to manage the system on a high level without overloadingit with demanding tasks. The key feature of accelerators isthat they are designed to handle tasks in parallel in a process called threading, making them more suitable for computationally demanding tasks. GPUs, FPGAs and ASICs are examples of suchhardware accelerators, and they are commonly used in MRI applications in processing complex image reconstruction and AI-based tasks [74]. GPUs excel in handling deep learning (DL) algorithms for image reconstruction, while FPGAs and ASICs offer customizable, power-efficient solutions tailored tospecific imaging needs [74]. For example, a recent study ona GPU-based pMRI has demonstrated improved processing speeds for image reconstruction [75]. Other studies focused on FPGA-based systems have been shown to optimize power consumption, which is an important aspect to improve the portability of pMRI systems [15]. ASICs have been designed to target specific aspects of MRI image processing, offeringa balance between speed and energy efficiency [76]. These developments highlight the growing importance of hardware acceleration in overcoming the computational and power limitations of pMRI systems. Below, we discuss each type of hardware accelerator.

### A. GRAPHICS PROCESSING UNITS (GPUs)

GPUs are parallel processing units originally designed for rendering graphics but have since evolved to handle complex computational tasks. They have been employed to accelerate the computationally intensive task of image processing, which relies on advanced algorithms to reconstruct a diagnosable image from raw data in MRI [77].

Figure 4 shows a simplified comparison of CPU and GPU architecture. As shown, GPUs' design trades more processing units (ALUs) for less memory cache capacity. Unlike CPUs, which are optimized for sequential processing and rely on a small number of powerful cores, GPUs consist of thousands of smaller, more specialized CUDAcores that excel at performing many operations in parallel. This gives them the ability to process multiple data streams simultaneously, making them well-suited for the high-speed, large-scale computations required in MRI [78].

GPU's utility is particularly evident in their superior performance with AI-based image reconstruction algorithms, as their architecture is specifically optimized to support operations such as convolutional and pooling layers through a combination of CUDA cores, tensor cores, and DL accelerators. Furthermore, theinherent parallelism of GPUs provides a significant advantage for iterative computations and matrix operations, as these taskscan be efficiently distributed and executed across thousandsof cores simultaneously, resulting in substantial reductions in processing time. In the context of pMRI, GPU hardware acceleration has been implemented either online using cloud GPUs or offline using edge GPUs. Cloud GPUs leverage thousands of CUDA cores optimized for floating-point 32-bit operations, delivering performance in the range of hundredsof Tera operations per second (TOPS). In contrast, edge GPUs incorporate fewer CUDA cores but include specialized hardware accelerators tailored for CNN computations (e.g.,DL Accelerator, Programmable Vision accelerator (PVA), and tensor cores), achieving efficient performance of a few hundred TOPS while operating at approximately 100 W ofpower consumption. In summary, cloud GPUs offer significantly higher computational power with scalability that isindependent of the pMRI system's physical size, enablingexpansion based on demand. However, their performance is heavily reliant on network connectivity, which can introduce limitations.

On the other hand, edge GPUs deliver lower overall computational power but can achieve comparable efficiencyfor specific applications due to their specialized architecture. Edge GPUs also consume less power compared to cloud GPU servers and operate independently of internet connectivity, ensuring reliable and consistent on-site performance, particularlyin remote or challenging environments.

Nvidia enhanced its GPU microarchitecture to tightly integrate tensor cores, which are specialized units for matrix operations targeting DL workloads. Such units offer high tensor arithmetic throughput, resulting in a substantial increase in the GPU's peak tera operations per second (TOPS). For example, NVIDIA T4 tensor cores provide up to 130 INT8 TOPS (130 TOPS for INT8) [114].

Very recently, one of the main manufacturers of brain p-MRI, Hyperfine Inc., with a market capitalization of $67 million, has partnered with the giant NVIDIA company [115]. The aim is to accelerate its AI-powered image reconstruction process without altering the image quality using NVIDIA's DALI and MONAI software tools.

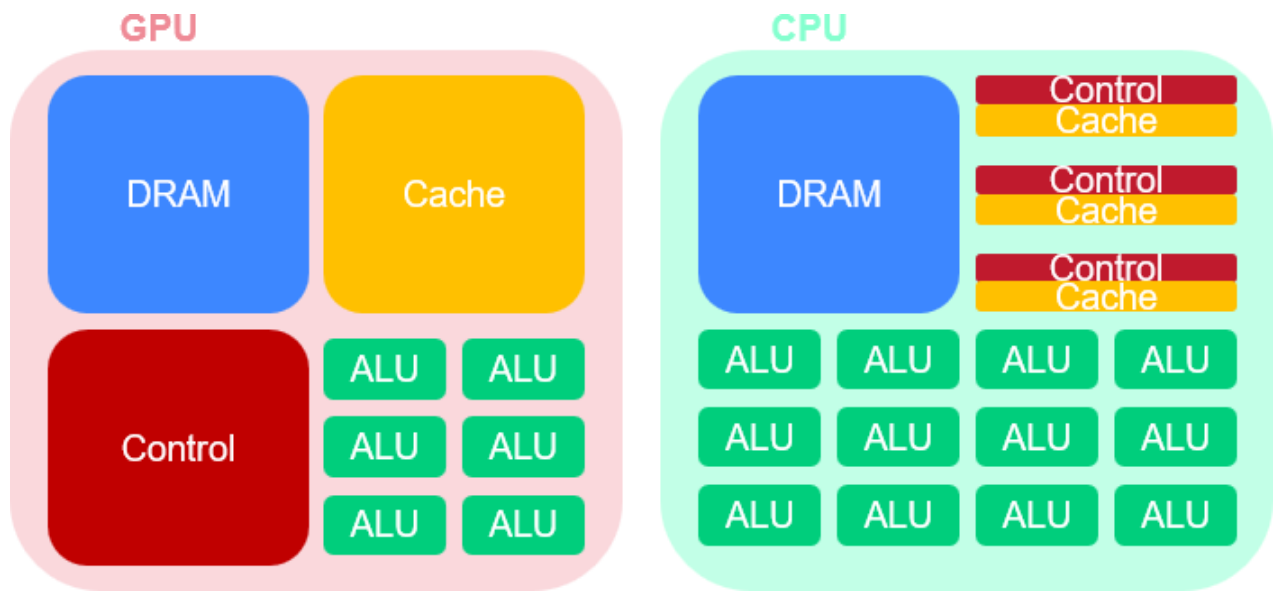

**FIGURE 4. Comparison of CPU and GPU Architectures [77]. ALU, arithmetic logical unit; GPU, graphics processing unit; DRAM, dynamic random access memory.**

### B. FIELD-PROGRAMMABLE GATE ARRAYS (FPGAs)

Field Programmable Gate Arrays (FPGAs) are reconfigurable hardware units initially developed for prototyping and digital circuit design. They have evolved into versatile platforms capable of accelerating complex computations and expediting the demanding task of image reconstruction, including real-time applications [79][80]. FPGAs' key advantage lies intheir customizable parallel architecture and low-latency processing capabilities. They enable application-specific hardwaredesigns that optimize data flow and minimize unnecessarycomputations. Unlike GPUs, which rely on fixed architectures and shared memory, FPGAs can be configured to implement tailored pipelines that efficiently handle the specific matrix operations involved in MRI reconstruction. This customizability allows for the optimization of memory access patternsand the reduction of processing delays.

Implementing FPGA-based acceleration in MRI reconstruction has been shown to significantly improve computation times while reducing powerconsumption [81]. For instance, A photoacoustic tomography experiment was conducted using a Xilinx Zynq-7000 (200 MHz Clock, 53200 LUTs, 4.9 Mb RAM) to reconstruct a 512x512 image using a delay-and-sum (DAS) reconstruction algorithm. Although the application is different, this algorithm shares a similar model to MRI parallel imaging algorithms. Compared to an i7-1165G7 CPU, which had a computational time cost of 1.344637s per image (22.31 FPS), the FPGA reduced the computational time to 0.021s per image (1,428.57 FPS). The FPGA-accelerated method achieved image reconstruction speeds 20 to 60 times faster than a CPU, with power consumption reduced from approximately 28 W to 1–2 watts [79]. Another comparative study evaluated the performance ofa proposed Xilinx Virtex-6 ML605 evaluation kit against an NVIDIA GeForce GTX 780 GPU and a Core i7 (2.9 GHz,4 GB RAM) CPU. The tests focused on a parallel imaging algorithm, SENSE (discussed in Section III), for reconstructing 256×256 images [81]. The results demonstrated a speedup of ×286.59 over the CPU and ×28.66 over the GPU, with a comparable signal-to-noise ratio (SNR) of approximately 35. To achieve these results, the FPGA utilized a pipeline structure, breaking down the matrix computations into smallerstages. This is useful for calculating the inverse of the matrix, $S^H S$ , which is critical for SENSE reconstruction. This was done using the adjoint and determinant method, expressed as:

$$(S^H S)^{-1} = \frac{adj(S^H S)}{det(S^H S)} \qquad (11)$$

where $adj(S^H S)$ is the adjoint (cofactor transpose) of the matrix and $det(S^H S)$ is its determinant.

To optimize computational efficiency, the multiplicative inverse of the determinant was precomputed and multiplied with the adjoint, eliminating the need for costly direct division operations. In [82], the FPGA-based implementation of real-time SENSE reconstruction is presented. The proposed system utilizes a parameterized pipeline architecture that efficiently performs the key steps of SENSE reconstruction, including matrix transposition, matrix multiplication, square matrix inversion, and magnitude computation. Figure 5 illustrates the data flow and hardware components of the system. The aliased image to be solved had a dimension size of 128 × 256 × 8. The implementation demonstrated a utilization of 729 DSP blocks (94%), 74,769 LUTs (49%), and 15 registers. The proposed FPGAs' exceptional performance stems from their ability to efficiently handle matrix operations, which are central to numerical reconstruction algorithms, as was discussed in the previous section. In contrast, DL accelerator FPGA units, while achieving up to 36.1× speedup over general-purpose CPUs, face challenges with scalability for large-scale networksdue to FPGA resource constraints and high memory demands [83]. While FPGAs can accelerate specific portions of neural network inference, such as tiled matrix multiplications, the iterative nature and large memory bandwidth requirementsin DL often favor GPUs. Thus, FPGAs are uniquely advantageous for deterministic, low-latency tasks like parallel imaging reconstruction, where operations can be pipelined and parallelized, outperforming CPUs and GPUs in power efficiency and consistent processing speed. For DL tasks, their performance remains limited to quantized inference.

For FPGAs, several proposals to improve the peak device throughput have coarsely integrated an FPGA fabric with a separate AI-optimized compute complex, such as in the Xilinx Versal architecture [116] or AI-targeted chiplets in Intel's system-inpackage ecosystem [117]. More recently, Intel introduced its first AI-optimized FPGA, the Stratix 10 NX, which integrates new AI tensor blocks and delivers up to 143 int8 and block fp16 TOPS [118]; a comparable peak performance to similar-generation GPUs.

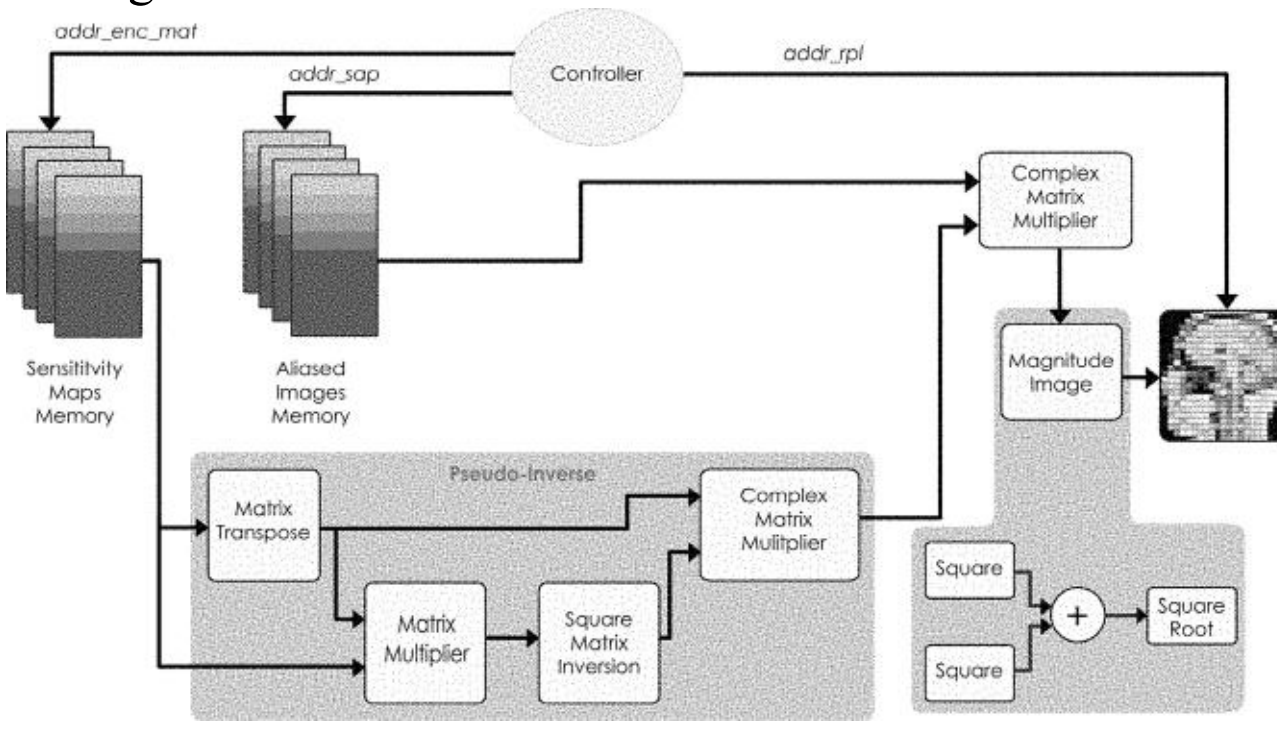

**FIGURE 5. Block diagram illustrating the data flow for MRI image reconstruction using SENSE. The system includes sensitivity maps and aliased image memories, a controller for managing data processing, matrix operations (transpose, multiplication, inversion) for pseudo-inverse computation, and a complex matrix multiplier for combining inputs. The final step computes the magnitude image for visualization, integrating square and. square root calculations [82].**

### *C. APPLICATION-SPECIFIC INTEGRATED CIRCUITS (ASICs)*

Application-Specific Integrated Circuits (ASICs) are custom-designed hardware units optimized for specific tasks, offering exceptional performance and energy efficiency. Unlike FPGAs, which are reconfigurable, ASICs are hardwired to perform predefined operations, eliminating reconfiguration overhead. This makes them ideal for computationally intensive, repetitive or deterministic tasks like image reconstruction in medical imaging, where speedand power efficiency are critical.

A widely recognized success story of ASIC technology is Google's Tensor Processing Unit (TPU), specifically designed to accelerate DL workloads. Unlike general-purpose GPUs, TPUs implement fixed-function hardware optimized for matrix multiplications and quantized operations essential for neural network inference. In tasks such as convolutional neural networks (CNNs), TPUs have been shown to deliver 15–30× performance-per-watt improvements over GPUs and CPUs. This success demonstrates the potential of ASICs to achieve unparalleled efficiency in repetitive, highly structured computations [84] [85]. In another study, an ASIC-based solution called Jigsaw was developed to accelerate non-uniform Fast Fourier Transform (NUFFT) (Discussed in Section III) computations in iterative MRI reconstruction [86]. The Jigsaw ASIC was specifically designed to handle the irregular sampling patterns common in MRI, which are computationally expensive when executed on CPUs or GPUs. To achieve this, the design incorporated pipelined and parallelized arithmetic units that efficiently handled the NUFFT kernels and reduced overall data movement. By implementing the NUFFT calculations in dedicated hardware, Jigsaw achieved gridding speedups of over 1500× compared to CPU implementations and 36× over state-of-the-art GPU implementations, demonstrating the immense computational advantage of ASICs in specialized workloads. Additionally, the ASIC reduced energy consumption by leveraging optimized fixed-point arithmetic and eliminating unnecessary data transfers, which are significant bottlenecks in GPU- and CPU-based systems. hardware solutions when considering rapidly evolving imaging algorithms. What makes ASICs unique in these implementations is their ability to achieve extreme levels of optimization for specific computational kernels, such as FFT, NUFFT, and matrix multiplications,by directly mapping these operations into hardware [87].

Unlike FPGAs, which rely on reconfigurable logic blocks, ASICs employ custom-built circuits that are tailored to the exact requirements of the algorithm. However, their fixed architecture poses challenges in adaptability. Unlike FPGAs, which can be reprogrammed for evolving algorithms, any modifications to an ASIC design require a costly and time-consuming hardware redesign process [88]. Nevertheless, for tasks such as Fourier-based image reconstruction and inference CNN models, where the computational model is well-defined and consistent, ASICs provide unparalleled performance and efficiency.

More recently, many different AI ASICs have been announced, such as Groq's Tensor Streaming Processors [119] and Graphcore's Intelligence Processing Unit [120.], that promise even higher peak performance of up to 820 int8 TOPS [121].

### D. COMPARATIVE ANALYSIS OF HARDWARE ACCELERATORS FOR pMRI

Table II illustrates the comparison of hardware accelerators including GPU, FPGA and ASIC based on latency, power, programmability, and system complexity. GPUs remain the most practical for algorithm development and rapid prototyping due to their mature software ecosystem and programmability. However, their large power draw and cooling requirements constrain their integration in battery-powered or portable MRI systems. FPGAs, in contrast, offer a favorable balance of low-latency execution, energy efficiency, and reconfigurability, making them the most viable hardware accelerator for portable deployments. Their ability to implement parallel pipelines for algorithms such as SENSE, GRAPPA, and CNNs has already been demonstrated with speedups of up to 298× over CPU baselines while consuming an order of magnitude less power [15]. ASICs achieve the highest efficiency and throughput, but their lack of flexibility and prohibitive design cost confine them to niche commercial systems where large-scale production justifies the investment.

At the algorithmic level, hardware suitability also varies. CNNs map naturally onto GPUs and FPGAs due to their regular convolutional structure, while GANs pose hardware inefficiencies because of transposed convolutions, requiring specialized optimizations [68]. Transformer-based models, which emphasize memory bandwidth and attention operations, may ultimately benefit from ASIC-like designs or carefully pipelined FPGA implementations.

In summary, GPUs dominate the research space, FPGAs enable practical portable systems, and ASICs remain a long-term path for ultra-low-power commercial imaging. These trade-offs provide a decision framework for selecting the appropriate accelerator depending on whether the application prioritizes flexibility, efficiency, or scalability.

## V. PORTABLE MRI SYSTEMS & ASSOCIATED HARDWARE ACCELERATORS

A comprehensive review of recent scientific literature was conducted to assess the state-of-the-art in pMRI technologies. Upon surveying published work on pMRIs, it was noted that there was significantly less focus on the back-end aspect compared to the front-end. Table III offers a comparison of the key article's results in a tabular format. The most notable pMRI applications will be discussed in this section.

### A. LIGHT PORTABLE BRAIN SCANNER

A pMRI prototype, developed by Massachusetts General Hospital and Harvard Medical School, utilizes a Halbach permanent magnet array to generate an 80 mT static magnetic field ($B_0$). This configuration eliminates the need for cryogenic cooling and high-power gradient systems, facilitating a more compact and energy-efficient design. The magnet itself weighs 122 kg, while the entire system, including coils and associated electronics, weighs approximately 230 kg. The system achieves a spatial resolution of 2.2 × 1.3 × 6.8 mm³ for a 20 cm diameter sphere target volume. Figure 6 depicts the design of the prototype. Among the pMRI applications reviewed in Table II, this prototype exhibited the lightest weight. Furthermore, the authors project that if the general-purpose prototyping components used, such as amplifiers, console, and cart, are replaced with custom lightweight designs, the overall weight can be reduced further from 230 kg to approximately 160 kg. A key method to highlight is the use of a built-in field gradient, which is integrated into the magnet design to perform spatial encoding. Traditionally, MRIs use magnetic coils to produce a linear gradient magnetic field for spatial encoding [90]. These gradient coils require high power supplies and equipment and contribute significantly to the heavy weights of conventional MRIs [91].

**TABLE II: Comparison of GPU, FPGA and ASIC**

| Accelerator | Speed (Latency/Throughput) | Power Efficiency | Flexibility/ Programmability | Design Complexity & Cost | Suitability for Portable MRI |
|---|---|---|---|---|---|
| **GPU** | High throughput (tens of ms to s); suited for large batch workloads | Moderate (10–200 W typical) | Very high (CUDA, PyTorch, TensorFlow) | Low design effort; hardware cost high | Limited (power and cooling needs restrict mobility) |
| **FPGA** | Low latency (µs–ms scale); demonstrated up to 298× CPU speedups | High (1–30 W typical) | Medium (HLS/RTL required; reprogrammable) | High design effort; moderate cost | Excellent (compact, energy-efficient, customizable) |
| **ASIC** | Ultra-low latency (ns–µs scale); fixed pipelines | Highest (sub-W per operation possible) | Very low (application-specific, not reprogrammable) | Very high non-recurring engineering (NRE) cost; long design cycle | Potentially ideal for high-volume commercial systems, but not flexible for research |

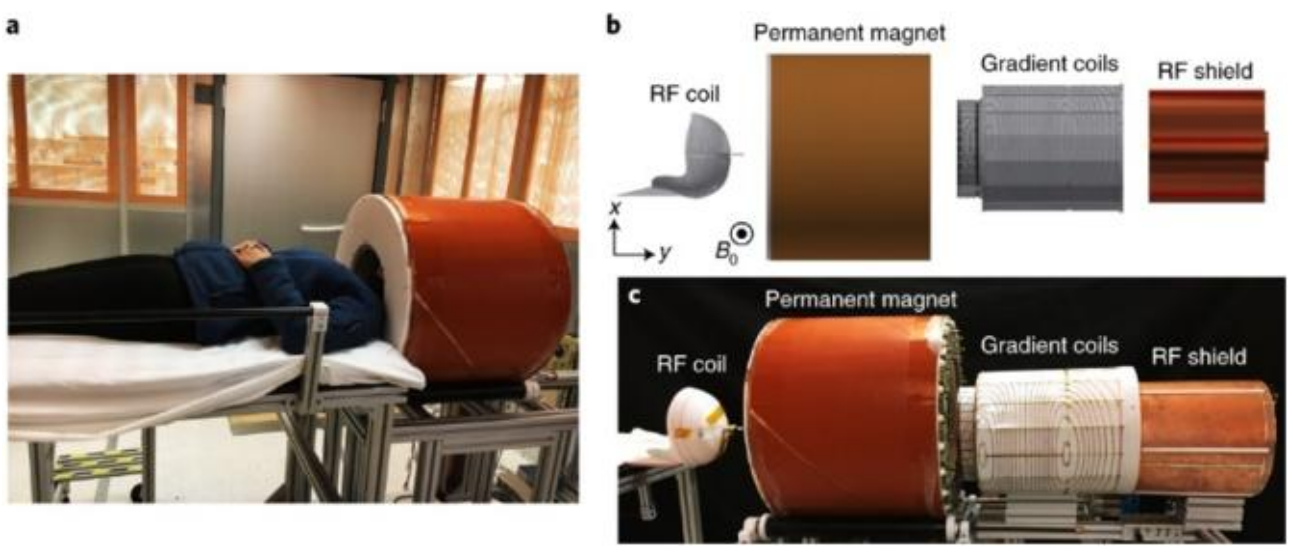

**FIGURE 6. Portable Brain Scanner Prototype by Cooley et al. Image used from the following cited source [89]**

The portable brain scanner reduces the reliance on coils by implementing an inherent gradient in the static magnetic field design. In other words, instead of traditionally superimposing a homogeneous field and a gradient magnetic field separately, the static magnet array was designed to provide a static field with an inherent gradient along the x-axis. Phase-encoding gradient coils are still utilized for encoding in the y and z directions. The noise introduced by this method is mitigated by implementing a model-based image reconstruction algorithm, specifically a conjugate gradient algorithm, which has distinct advantages for this application due to its ability to iteratively approximate solutions for large, sparse, and linear systems of equations. Compared to the conventional inverse Fourier transform-based image reconstruction, model-based image reconstruction algorithms are more computationally expensive [92]; however, they allow the incorporation of prior knowledge of the image into the reconstruction process. In the portable brain scanner, the authors included the inhomogeneity resulting from the permanent magnet array design into the iterative reconstruction process. Figure 7 illustrates the image quality of the portable brain scanner using T1 and T2 contrasts. The image reconstruction process was done online for these scans; however, the authors mention the applicability of hardware accelerators, such as the Tesla K20c (5 GB) and the Tesla P100 (16 GB), based on the size of their global shared memory. Overall, this work presents a perspective on how the portability of the MRI is directly related to the back-end electronics dedicated to image reconstruction. Although offloading the functionality of a readout gradient to the reconstruction software increases the overall scan time, it still provides an opportunity to build lighter pMRI devices [89].

### B. A FAST BRAIN SCANNER

Recent progress in portable low-field MRI has shown that clinically useful imaging can be achieved by combining lightweight hardware with computational acceleration. A representative example is a 55 mT ultralow-field (ULF) pMRI system that uses static magnetic field generated by a samarium-cobalt. permanent magnet.The system integrates a fast spin echo (FSE) sequence for T1- and T2-weighted brain imaging, achieving scan times of 2.5 and 3.2 minutes, respectively. The system weighs a net of 750 kg and is designed for brain imaging only. This system demonstrates the potential of DL models in pMRI applications. In 2023, the authors proposed a design that utilizes a CNN model to enhance the quality of low-field images through super resolution (SR) [12]. The model is trained on a high field (3 T) and publicly available dataset called the Washington University - Minnesota University Human Connectome Project [93]. The results show that the application of the CNN model improved the spatial resolution of the scanned images from 3 x 3 x 3 mm3 to 1.5 x 1.5 x 1.5 mm3.

Recently, the authors extended the scanner to use a single-average 3D encoding with 2D partial-Fourier (PF) sampling to shorten acquisitions, and a 3D deep-learning reconstruction (PF-SR) to recover image quality. The learning model leverages prior brain anatomy from high-field data to suppress noise and artifacts and to enhance spatial fidelity at low field. Inference is executed on a GPU, so that computational latency does not dominate the end-to-end exam time [75].

The model was trained on 4 NVIDIA V100 GPUs, and the inference was run on a single NVIDIA V100 GPU, which could reconstruct a single 160 × 192 × 160 voxel anatomical image in 1.9 seconds. This reduced the overall scan times of T1W & T2W sequences from 8.6 & 11.2 mins to 2.5 & 3.2 mins, respectively. The model demonstrated high accuracy, achieving consistently higher 3D Structural Similarity Index Measure (SSIM) and lower Normalized Root Mean Square Error (NRMSE) compared to traditional non-DL methods, as validated on 200 synthetic ULF datasets.

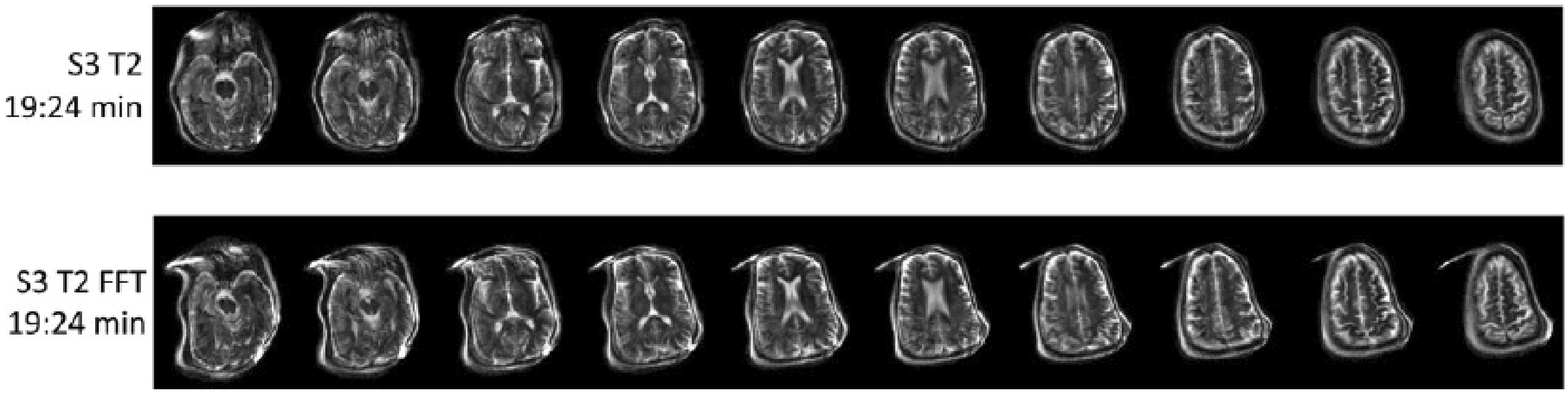

**FIGURE 7. Acquired images by the portable brain scanner. The acquisition time of the sequence is shown to the left. The top row utilizes model-based image reconstruction, while the bottom row depicts a conventional Fourier-transform-based reconstruction. The measured SNR in the top image is 124. [89]**

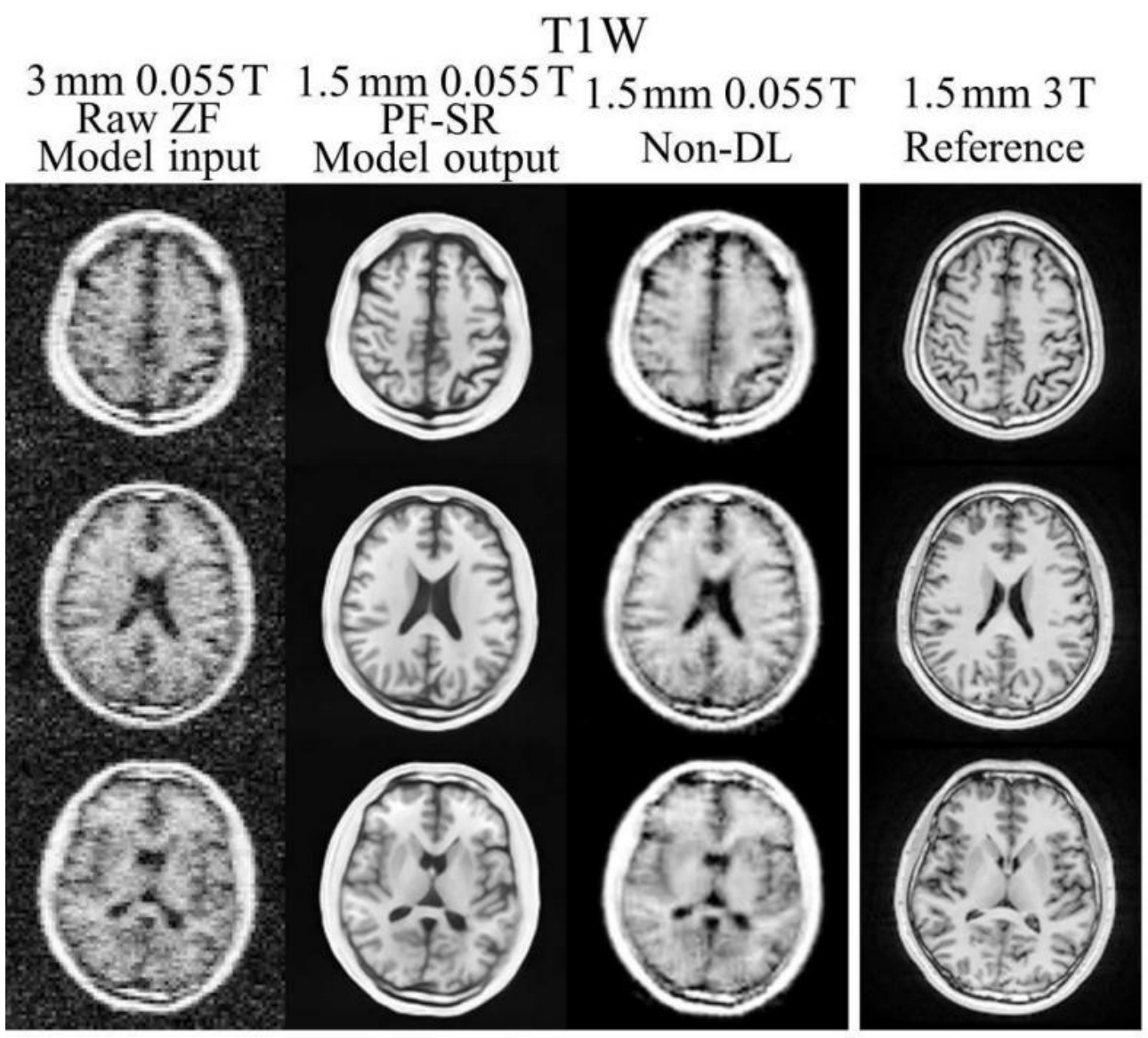

**FIGURE 8. Reconstruction of experimental low-resolution 3D brain data with prospective 2D PF sampling from one healthy volunteer (33-yearold male), acquired from a low-cost shielding-free 0.055-T MRI head scanner, using traditional non-DL and PF-SR. Images used from the following cited source [75].**

Figure 8 depicts a comparison between non-DL-based images and partial Fourier SR-based images of the same subjects. As demonstrated, the limitations of scanning speed and image quality in low-field MRIs can be overcome through the introduction of DL models, which opens new avenues for the development of high-performing low-field MRIs for the future. From a hardware-acceleration perspective, this case illustrates a practical pattern for portable MRI: simplify the magnet and encoding to fit point-of-care constraints, then recover image quality and resolution via GPU-accelerated reconstruction. The use of hardware acceleration, specifically the NVIDIA V100 GPU, was instrumental in enabling the DL model to minimize the reconstruction of high-resolution 3D anatomical images. This highlights the critical role of hardware acceleration in overcoming the computational challenges of DL-based reconstruction, making fast, high-quality imaging feasible even for low-field pMRI systems. This case illustrates that by shifting the performance bottleneck from acquisition to GPU-based reconstruction, fast and high-quality imaging becomes feasible in portable low-field systems.

### C. OPEN pMRI FOR IMAGING EXTREMITIES

A pMRI system for imaging extremities was developed by researchers at Université de Lorraine, Nancy, France, offering a significant leap in both portability and technological advancements [9]. This low-field MRI system generates a static magnetic field of 0.295 T using two axially magnetized Neodymium-Iron-Boron (NdFeB) permanent magnets. The open-field configuration is designed to provide ease of access and improved flexibility, positioning it as a viable alternative to traditional closed MRI systems, particularly in low-resource settings and mobile medical environments. The system's compact size (35 x 35 x 30 cm) and lightweight build (120 kg) enable it to be deployed in mobile medical units, field hospitals, or emergency care scenarios, where stationary MRI systems are impractical. Furthermore, its design eliminates the need for cryogenics or heavy infrastructure, addressing the operational limitations of conventional MRI scanners.

However, this pMRI system faces challenges at higher frequencies due to eddy current-induced gradient field attenuation in the conductive components (iron and magnets) due to its relatively high $B_0$. To address this, integrating hardware-accelerated solutions could significantly improve its performance. For example, a solution leveraging FPGA-based real-time compensation can be implemented. Based on a demonstration provided in [39], this approach would integrate an FPGA control loop to dynamically adjust the gradient coil currents based on real-time feedback. By monitoring field distortions using dedicated sensors, the FPGA can calculate and apply compensatory currents with low latency, effectively mitigating the unwanted eddy currents. This could enable the system to maintain its gradient strength of 8 mT/m even at higher frequencies, overcoming the limitations posed by eddy currents. The resulting improvements in performance would enhance the system's capability for real-time, high-quality imaging in critical care and remote settings. Although there is no explicit mention of hardware acceleration in this work, its potential remains valuable in further improving the innovative design of front-end-focused prototypes.

### D. AN FDA-CLEARED pMRI SYSTEM

The Swoop pMRI system (Hyperfine, Inc.) weighs 630 kg, with a height of 140 cm and a width of 86 cm. The product operates with a standard wall outlet, a power supply of 15 A, 110V, and a static magnetic field of 64 mT [4]. The Swoop implements DL model-based image reconstruction to achieve an increase of 60% in the SNR of T1W, T2W, and FLAIR scans [94]. As can be seen in Figure 9, the design is specialized for brain imaging.

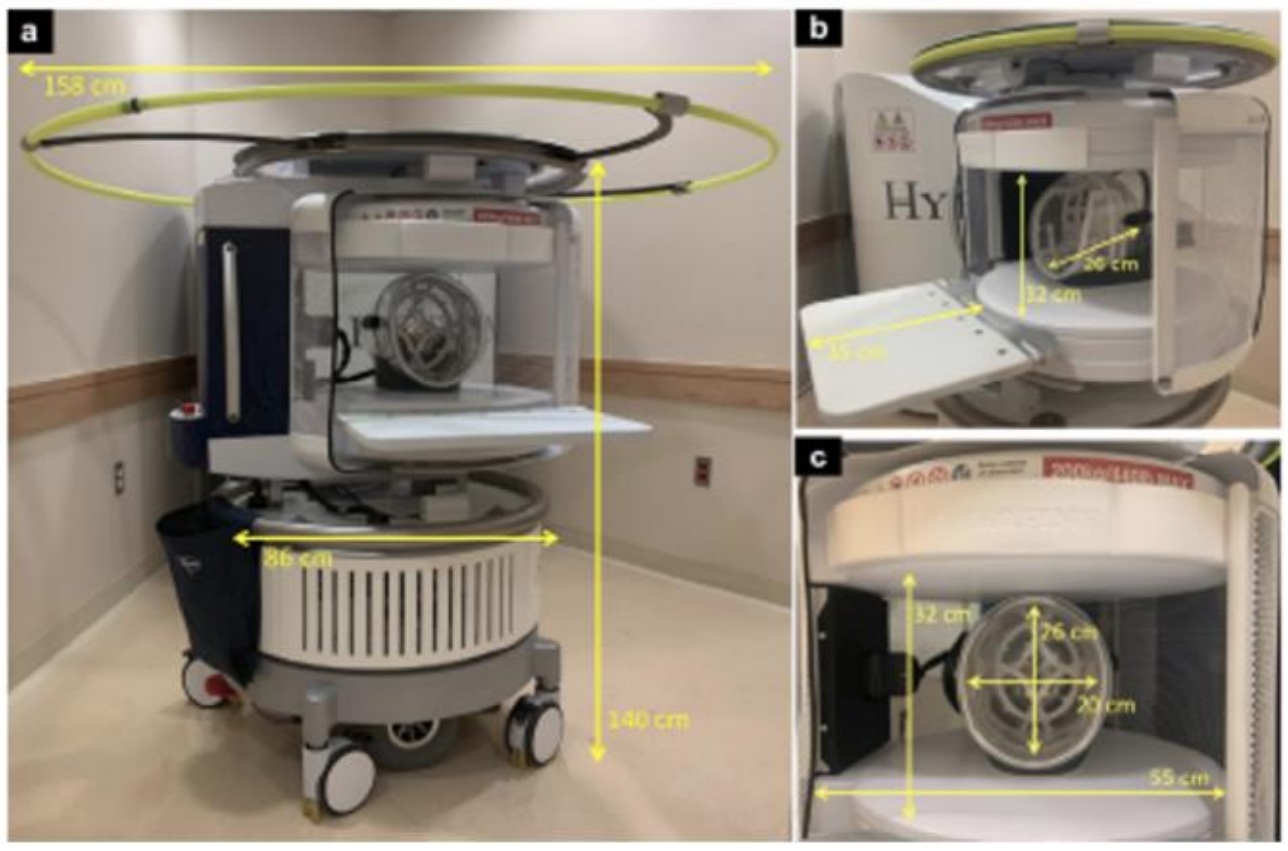

**FIGURE 9. Measurements of SwoopR pMRI system. The image is used from the following source, [4].**

As it is the only FDA-cleared model, this system provides a reference for the practicality of pMRIs in a commercial setting. This makes the system very attractive for MRI in low- and middle-income countries [95]. There is limited publicly available information about hardware accelerators in the Swoop Portable MRI System. Based on existing data, the Swoop system primarily utilizes cloud-based GPUs for preprocessing and post-processing tasks, specifically advanced gridding and denoising steps during image reconstruction. These tasks enhance image quality by reducing artifacts and noise. However, the reconstruction process itself does not appear to rely on onboard hardware accelerators. As a result, the system's performance may be highly dependent on the available network connectivity.

Table III shows clear trade-offs across existing pMRI implementations: Fast 3D brain MRI prioritizes acquisition speed (GPU + DL) and yields the shortest scan times, Open pMRI (highest B0 ≈ 295 mT) provides superior intrinsic SNR for small-target imaging, the Portable Brain Scanner offers the best whole-brain portability vs performance trade-off (lightest whole-brain design with GPU-accelerated iterative/DL reconstruction), and the Hyperfine Swoop represents the most clinically mature option (FDA clearance and bedside use) despite lower B0 and some limits in diagnostic sensitivity. These distinctions illustrate how choices of B0, target volume, accelerator hardware, and reconstruction algorithm map directly to different application niches (speed, SNR, portability, or clinical readiness).

**TABLE III: Comparison Of State-Of-The-Art Applications Of pMRIs**

| **pMRI** | | **Portable Brain Scanner** | **Fast 3D brain MRI** | **Open pMRI** | **Hyperfine Swoop** |
|---|---|---|---|---|---|
| **References** | | [89] | [75] | [9] | [4] |
| **$B_0$ (mT)** | | 80 | 55 | 295 | 64 |
| **$V_{Target}$ (DSV)** | | 200 mm | 240 mm | 100 mm | Not mentioned |
| **HW Accelerator** | | Tesla K20c/Tesla P100 (GPU) | NVIDIA V100 (GPU) | Not mentioned | Not mentioned |
| **Algorithm** | | Conjugate Gradient | DL | Not mentioned | DL |
| **Weight (kg)** | | 230 | 750 | 120 | 630 |
| **Sequence Acquisition Time (min:sec) [Spatial Resolution ($mm^3$)]** | **T1W** | 11:46[2.2 ×1.3×6.8] | 2:30 [15×1.5×1.5] | Not mentioned | 4:54 [1.5×1.5×1.5] |
| | **T2W** | 9:42 [2.2×1.3×6.8] | 3:12 [1.5×1.5×1.5] | Not mentioned | 7:03 [1.5×1.5×5] |
| | **FLAIR** | Not mentioned | Not mentioned | Not mentioned | 9:31 [1.5×1.5×5] |
| | **DWI** | Not mentioned | Not mentioned | Not mentioned | 9:04 [2×2×5] |
| | **PDW** | 9:24 [2.2×1.3×6.8] | Not mentioned | Not mentioned | Not mentioned |
| **Advantages** | | Enables clinically useful whole-brain imaging via GPU-accelerated iterative and DL reconstruction, improving low-field image quality and reducing end-to-end processing time for point-of-care use. | Very fast 3D acquisition and high spatial resolution; GPU enables rapid recon | High $B_0$ for small-target imaging (good SNR for extremities); open geometry possible | FDA-cleared, compact (<1.5 T weight), head imaging at bedside; GPU-accelerated reconstruction improves usability; enables neuroimaging in ICU/ emergency settings, clinically validated for stroke and pediatric imaging. |
| **Limitations** | | Longer acquisition times vs high-field; bulky due to GPU; moderate SNR | Heavy system (weight), high power consumption — less portable in practice | Small target volume only; specialized use (not whole-brain); HW accel. not reported | Lower SNR vs high-field, limited diagnostic sensitivity for some pathologies; longer scan times for certain sequences vs high-field |

## VI. CLINICAL VALIDATION AND REAL-WORLD DEPLOYMENT CHALLENGES

Despite rapid progress in hardware acceleration and AI based reconstruction, the clinical validation of pMRI systems remains limited compared to high-field MRI. Most published demonstrations have been pilot studies with relatively small cohorts, often focused on feasibility (e.g., brain imaging in low-resource environments) rather than full-scale clinical trials. For instance, Hyperfine's Swoop™ system has received FDA clearance for brain imaging at the point of care, but widespread clinical adoption is still constrained by the need for rigorous multicenter validation across diverse populations and pathologies [105]. Similarly, ultralow-field (55 mT) prototypes have shown promising results in stroke triage and pediatric neuroimaging, yet larger-scale clinical outcome studies are needed to confirm diagnostic equivalence to conventional systems. Without such validation, the integration of pMRI into standard care pathways will remain incremental.

Regulatory approval presents another major challenge. While AI-based reconstruction can be cleared as part of hardware systems under regulatory frameworks (e.g., FDA 510(k), CE marking), adaptive or continuously learning models complicate approval processes. Regulators increasingly require explainability, robustness testing, and post-market monitoring of AI algorithms to ensure safety. For hardware-accelerated systems, additional evaluation is necessary to demonstrate stability under diverse workload conditions and to validate real-time performance claims. Collaborative initiatives between vendors, regulators, and clinical institutions are therefore essential to accelerate safe approval pathways.

Interoperability and standardization also play a critical role in deployment. Current pMRI devices often operate with proprietary acquisition protocols and reconstruction pipelines, which hinders integration with existing Picture Archiving and Communication Systems (PACS) and hospital IT infrastructure. Adoption of standards such as DICOM for image exchange and HL7/FHIR for electronic health record (EHR) integration will be vital for seamless clinical workflows. Standardization would also facilitate benchmarking across studies, allowing more systematic comparisons of AI reconstruction methods and hardware platforms.

User experience and training requirements are another underexplored dimension. Unlike conventional MRI, which requires trained technologists and radiologists in specialized imaging suites, portable MRI is envisioned for bedside or remote operation by non-experts. This shift necessitates simplified interfaces, automated calibration, and AI-driven acquisition protocols to reduce operator dependency. Initial deployments in critical care units suggest that non-specialists can acquire clinically useful scans with minimal training, but further research is needed to optimize workflows, reduce operator errors, and ensure consistent image quality. Human factors engineering such as ergonomic design, intuitive software, and decision support tools, will be critical to enabling adoption in real-world clinical environments.

Reimbursement and economic considerations must not be overlooked. For pMRI to be sustainable, clear cost-effectiveness data are needed, demonstrating that accelerated hardware platforms can reduce imaging bottlenecks, shorten hospital stays, or expand access in underserved regions. Early health-economic studies suggest potential savings in stroke triage and neonatal imaging, but systematic analyses are limited. Establishing robust clinical and economic value propositions will ultimately drive adoption beyond early research and pilot projects.

## VII. DISCUSSION AND FUTURE PROSPECTS

The future of pMRI systems lies in overcoming the inherent challenges posed by the use of low-field systems ($B_0$ lower than 0.1 T), including low signal-to-noise ratio (SNR) and prolonged scan times. Parallel imaging techniques, traditionally limited by SNR at low field strengths, are now being revitalized with AI-based methods that leverage the power of DL to boost image reconstruction speed and quality. For example, Deep De-Aliasing Generative Adversarial Networks (DAGAN) have shown the ability to reconstruct high-fidelity images from highly undersampled k-space data, achieving up to 10× acceleration while maintaining PSNR values above 31 dB [96]. These AI models excel at suppressing aliasing artifacts and enhancing fine details, compensating for the low SNR at low magnetic fields. Similarly, AI has demonstrated improved stability and robustness in highly undersampled and noisy environments [97]. By combining these AI methods with parallel imaging, pMR systems have the potential to achieve clinically viable scan speeds without compromising image quality. Hyperfine's recent work with self-supervised learning methods highlights the potential of AI in low-field MRI applications [98].

Despite this progress, one of the major challenges in applying AI to pMR is the lack of publicly available low-field MRI datasets. As demonstrated in the earlier section, most AI-based reconstruction models are trained on high-field MRI datasets, which do not fully capture the noise characteristics, artifacts, and resolution constraints of low-field portable systems. This mismatch poses a significant limitation for generalizability [19]. Without sufficiently diverse low-field-specific datasets, AI models may fail to deliver optimal performance in real-world pMR applications. Transfer learning approaches, where models pretrained on high-field datasets are fine-tuned for low-field data, offer an alternative solution. Overall, the creation of dedicated public datasets for low-field MRI remains essential for advancing AI robustness and clinical adoption in portable settings.

To move beyond problem identification, we propose a concrete, actionable call to action: the rapid formation of a multi-institutional Low-Field MRI Consortium tasked with curating, standardizing, and openly sharing low-field datasets and benchmarks. The consortium should commit to delivering:

(1) A publicly accessible "Low-Field MRI Dataset v1.0" (pilot goal: ~200 de-identified scans from ≥3 sites

covering common pulse sequences and clinical indications).

(2) A mandatory metadata schema (B0, coil configuration, scanner model, pulse sequence parameters, k-space sampling pattern, SNR estimates, acquisition geometry, anonymized clinical labels, and provenance).

(3) A set of benchmark tasks and metrics (reconstruction: acceleration factors with PSNR/SSIM and task-aware scores; downstream tasks: segmentation Dice, diagnostic concordance vs high-field) and an online leaderboard.

(4) Validated synthetic-data pipelines (Bloch-based simulators and GAN/physics-informed methods) with documented generation parameters. Governance should support both centralized sharing and privacy-preserving distributed modes (federated learning, tiered data use agreements), and include a steering committee representing industry, clinical sites, and academic partners.

To bootstrap adoption, we recommend a 12–18-month pilot funded by a mixed public–private mechanism that releases a baseline dataset, containerized baseline algorithms, and reproducible evaluation scripts. By defining deliverables, metadata, evaluation criteria, and governance up front, this consortium would make low-field MRI datasets discoverable, comparable, and clinically meaningful, accelerating reproducible AI research and translation for portable systems.

Beyond expressing the need for real low-field datasets, emerging work also explores how to generate such data synthetically. For example, denoising Cycle-GANs have been used to convert simulated low-field (e.g., 70 mT) images, created by injecting Rician noise into high-field data, into high-SNR outputs with good fidelity, evaluated using SSIM and PSNR metrics [101]. More broadly, GANs and VAEs have seen widespread application in augmenting brain MRI datasets, by capturing realistic anatomical distributions suitable for downstream tasks like segmentation [102], [103]. An especially promising approach is the use of Bloch equation–based simulators, which integrate physical magnetization models over k-space trajectories to generate synthetic MRI data with controllable inhomogeneities and noise characteristics [104]. By combining such physics-informed simulations with GAN-based image synthesis, one can create diverse and realistic low-field datasets for improved AI model training, enabling systematic benchmarking under controlled conditions.

The hardware infrastructure of pMR systems is another critical consideration for enabling improved and reliable AI implementation. The Swoop pMR likely relies on cloud-based GPUs for processing, which introduces dependencies on network connectivity and increases latency risks in remote or low-resource environments. In contrast, the adoption of Edge GPUs offers a more reliable and efficient solution. Edge GPUs eliminates the reliance on Internet access, providing consistent performance in isolated clinical settings or remote environments. A particularly exciting opportunity lies in the design of lighter, more portable MRI systems that intentionally allow increased noise but at simplified hardware designs. While hardware simplifications would result in lower SNR and degraded image quality, AI-based reconstruction techniques can compensate for these limitations by effectively denoising and correcting for hardware-induced imperfections. This can be seen in [89] [99], where iterative reconstruction methods successfully alleviated the constraints of lightweight hardware designs.

However, shifting diagnostic fidelity from the hardware toward AI reconstruction also introduces unique validation challenges: how to demonstrate equivalence when diagnostic quality depends more on software reconstruction than on the acquisition hardware itself. In such cases, substantial equivalence must be argued at the system level, showing that the combined hardware–AI pipeline produces diagnostic outcomes on par with conventional MRI. This requires new evidence strategies, such as benchmarking diagnostic concordance against high-field systems, evaluating robustness under noisy acquisition conditions, and validating AI across diverse populations. Rather than treating hardware and AI separately, regulators may increasingly demand integrated frameworks that jointly assess acquisition fidelity and reconstruction reliability. Developing such validation methodologies will be key to enabling regulatory approval of AI-compensated lightweight pMRI systems.

To operationalize the regulatory requirements described above, we propose a clear evidence ladder that stages validation from controlled technical tests to full clinical evaluation. The following three steps form the core progression investigators should follow to demonstrate safety and diagnostic equivalence for AI-reconstructed images from lightweight pMRI systems:

- Analytic / phantom validation: standardized phantoms and controlled k-space injections to quantify SNR, spatial fidelity, artifact response, and out-of-distribution failure modes.
- Retrospective multi-center testing: evaluation on curated low-field datasets (pilot goal: ~200 de-identified cases from ≥3 sites) measuring both image-quality metrics and task-aware clinical endpoints.
- Prospective reader studies / non-inferiority trials: multi-site reader studies followed, where indicated, by powered non-inferiority clinical trials to demonstrate diagnostic concordance with high-field MRI for prespecified endpoints.

Crucially, evaluation metrics must extend beyond PSNR and SSIM to include sensitivity, specificity, diagnostic concordance, and inter-reader variability, reflecting outcomes meaningful to clinicians and regulators.

Regulatory pathways will depend on system design: locked AI models paired with fixed lightweight hardware may be eligible for 510(k) or De Novo submissions, whereas adaptive or continuously learning systems are more likely to require a Premarket Approval (PMA) process with ongoing performance monitoring [106]. In all cases, lifecycle controls, including version management, change protocols, and cybersecurity safeguards, will be essential to maintain safety and compliance. Adherence to international standards such as IEC 62304 [107] (software lifecycle), ISO 14971 [108] (risk management), and ISO 13485 [109] (quality management) will provide a recognized framework for device approval.

The proposed Low-Field MRI Consortium could also extend its role to regulatory alignment by releasing standardized datasets, containerized baseline algorithms, and reproducible evaluation scripts that regulators may reference as testbeds. This dual function, supporting both research reproducibility and regulatory validation, would accelerate the safe translation of AI-compensated lightweight pMRI systems into clinical use.

The selection of GPUs, FPGAs, or ASICs in future pMRI systems is less about raw performance and more about the trade-offs between flexibility, latency, and power efficiency. For teams focused on rapid prototyping and evolving AI models, edge GPUs remain the most practical option given their mature software ecosystems and flexibility, even if they incur higher power costs. In contrast, applications with fixed, latency-critical pipelines, for example, real-time stroke detection using a stable, regulatory-approved algorithm, may ultimately benefit from ASIC implementations, despite the high non-recurring engineering (NRE) costs. FPGAs occupy a middle ground: their reconfigurability makes them well-suited for real-time preprocessing tasks such as shimming, gradient control, or on-the-fly data compression, where deterministic timing is critical, while they can operate in tandem with GPUs for heavier reconstruction workloads. In practice, future pMRI systems will likely adopt heterogeneous architectures, where lightweight acquisition hardware is paired with hybrid accelerators, enabling a balance of adaptability, energy efficiency, and clinical robustness.

Looking forward, several emerging technologies can further transform pMRI. Quantum computing offers promise for solving large-scale inverse problems in MRI, potentially enabling real-time optimization of reconstruction and pulse-sequence design. Neuromorphic chips, with their event-driven and massively parallel architecture, present a path toward ultra-low-power AI inference, which is particularly relevant for portable systems operating in resource-constrained environments. Together, these computing paradigms could augment existing FPGA and GPU accelerators, providing new levels of efficiency and scalability.

Another frontier is the integration of pMRI into edge–IoT healthcare ecosystems. Future systems may leverage IoT connectivity for seamless integration with telemedicine platforms, remote monitoring, and hospital networks, allowing scans to be initiated, reconstructed, and reviewed across distributed sites. This opens opportunities for population-scale screening, especially in rural or underserved regions. However, increased connectivity also raises cybersecurity risks, including patient data breaches, manipulation of raw k-space data, or adversarial attacks on AI reconstruction models. End-to-end encryption, secure firmware updates, and anomaly detection in reconstruction pipelines will be crucial for safe deployment.

sustainability and environmental considerations are increasingly important. Portable low-field MRI already eliminates reliance on cryogenic cooling and helium, reducing environmental impact relative to high-field superconducting systems. However, the energy demands of GPUs and large AI accelerators present a new challenge. Optimizing pipelines for FPGA or neuromorphic accelerators, prioritizing energy-efficient model architectures, and designing recyclable permanent-magnet systems are critical to ensuring that future pMRI solutions are not only clinically effective but also environmentally sustainable.

The convergence of AI-based reconstruction, edge hardware acceleration, emerging computing paradigms, and sustainable design is reshaping the landscape of portable MRI. By addressing data limitations, ensuring cybersecurity, and embedding pMRI within broader healthcare ecosystems, future systems can deliver fast, high-quality imaging at the point of care. The continued co-design of lightweight hardware with advanced accelerators will be central to enabling the next generation of reliable, accessible, and sustainable portable MRI.

## VIII. CONCLUSION

The development of pMRI systems presents a transformative opportunity for accessible medical imaging, particularly in remote and resource-limited settings [100]. This paper highlights the pivotal role of hardware acceleration in addressing the computational challenges associated with pMRI, including image reconstruction speed, energy efficiency, and realtime processing. We have surveyed how Edge GPUs, FPGAs, and ASICs enable execution of complex reconstruction algorithms and AI models with reduced latency and power. The following are the key findings through the survey:

- Hardware acceleration is essential for pMRI viability. Edge GPUs, FPGAs, and ASICs provide the throughput and energy efficiency required to run advanced reconstruction and ML models on- or near-device, making real-time or near-real-time pMRI workflows feasible.
- Edge processing reduces dependence on network connectivity. Compared with cloud-based pipelines, edge acceleration offers more reliable and predictable performance in remote or low-resource environments where connectivity is unreliable.
- AI-based reconstruction complements hardware advances. Generative models and DL denoising substantially improve image quality at low field strengths, enabling aggressive undersampling and shorter acquisitions when supported by sufficient compute.

- Algorithm–hardware co-design is a promising path. Techniques such as model-based reconstruction, parallel imaging adapted for low-field, and quantized/compressed DL models are most effective when jointly designed with the target accelerator (GPU / FPGA / ASIC).
- Trade-offs remain between performance, power, and flexibility. GPUs offer rapid development and strong DL support but consume more power; FPGAs provide deterministic, low-power operation with higher development cost; ASICs maximize efficiency but are costly and inflexible to update.
- Data availability is a major bottleneck for generalizable AI. Public, diverse low-field pMRI datasets are essential to train and validate AI reconstructions that generalize across devices and populations.
- Practical deployments are emerging but still limited. Commercial systems (e.g., Hyperfine Swoop) and multiple research prototypes show feasibility and clinical utility, yet the back-end acceleration ecosystem for pMRI remains nascent and ripe for innovation.

Moving forward, the integration of hardware-accelerated AI methods, along with efforts to develop low-field MRI datasets, will be critical to achieving the full potential of pMRI technology. Future work should emphasize algorithm–hardware co-design, low-power inference strategies (quantization, pruning), and standardized benchmarks for pMRI reconstruction and clinical performance. These advancements will drive the next generation of pMRI systems, delivering high-quality imaging at reduced scan times while maintaining portability and energy efficiency. Hardware acceleration thus remains at the heart of enabling real-time, portable, and scalable MRI solutions capable of bridging the inequality gap in global healthcare accessibility.

## ACKNOWLEDGMENT

The authors would like to acknowledge Khalifa University of Science & Technology (KUST) of Science & Technology and the KUST's HEIG group for supporting this work.

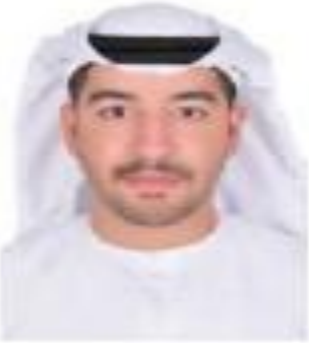

**Omar Alhabshi** is a professional engineer at ASEA Brown Boveri (ABB) Transmission & Distribution Ltd., specializing in process automation systems. He received a B.S. in Electrical Engineering from Colorado State University in 2022. During his final year, he developed operational risk solutions on the Petawatt-Class Advanced Laser for Extreme Photonics (ALEPH) at the Advanced Beam Laboratory, CSU . He has 2 years of experience in designing and implementing advanced automation systems.

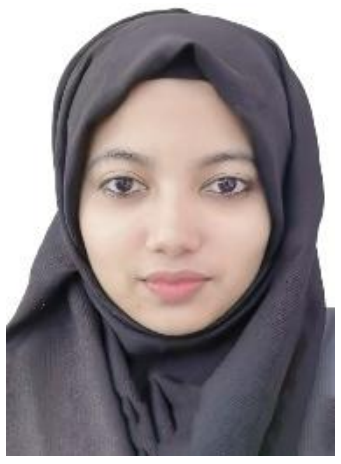

**Safa Mohammed Sali** received B.Tech in Electronics and Communication Engineering and M.Tech in Embedded Systems from APJ Abdul Kalam Technological University, Kerala, India. She is currently a visiting researcher in Computer and Information Engineering at Khalifa University of Science and Technology, Abu Dhabi, United Arab Emirates.

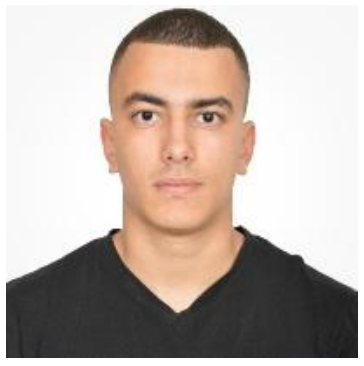

**Anis Meribout** MD, is a physician-scientist at the Biomedical Engineering and Imaging Institute, Icahn School of Medicine at Mount Sinai. He earned his medical degree with high honors from the University of Constantine 3 – Salah Boubnider, Algeria, in July 2024. His research focuses on liver and abdominal MRI, with emphasis on radiomics, quantitative imaging biomarkers, and artificial intelligence in oncology. He has led MRI-based radiomics projects in hepatocellular carcinoma and published on the applications of large language models in medical tasks and cancer imaging.

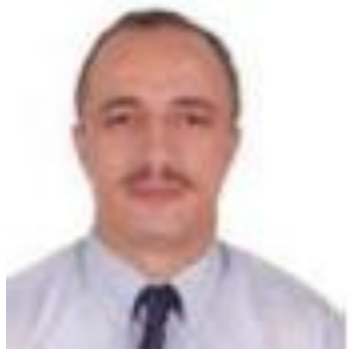

**Prof. Mahmoud Meribout** (M'91) received a B.Eng. degree in 1985 and a Ph.D. degree in electronics engineering from the University of Technology of Compiegne, Compiegne, France, in January 1995. He was with Nippon Telegraph and Telephone Corporation, Tokyo, Japan, and NEC Corporation, Tokyo, from 1995 to 2000, where he was involved in several projects related to embedded systems design. In 1998, he received the NTT Best Award for his research and development efforts in the areas of embedded systems design and imaging systems. In 2008, he joined the Electrical Engineering Department, Petroleum Institute (now Khalifa University of Science & Technology, Abu Dhabi, UAE), where he is currently a full Professor. His current research interests focus on embedded systems, imaging systems (THz, MPI, optical, electrical, and magnetic tomography), instrumentation, and multiphase flow metering.

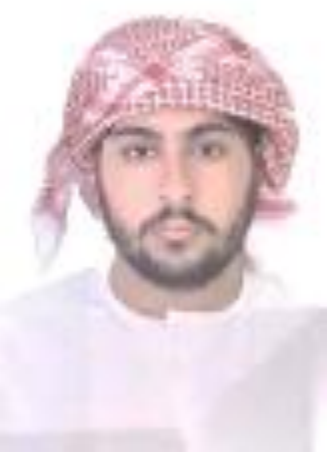

**Saif Almazrouei** is a researcher at the Directed Energy Research Center in the Technology Innovation Institute. He obtained a Bachelor of Science degree from the University of Penn State in Physics, a Bachelor of Arts degree from the University of Penn State in Philosophy in 2020, and a Master's degree in applied physics specializing in High Power Laser and Optics from the Belarusian State University in Minsk in 2023. He conducted research in the field of quantum imaging, investigating the utilization of Fisher information and Cramer-Rao bound in parameter estimation. He also conducted research in Raman spectroscopy and Interferometry.

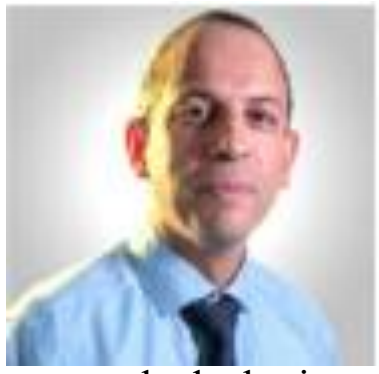

**Prof. Mohamed Lamine Seghier** is an expert in functional MRI and cognitive neuroscience. He earned his PhD from Grenoble's Joseph Fourier University and completed a post-doc at Geneva University Hospitals. He was a Senior Research Fellow at the Wellcome Centre for Human Neuroimaging, UCL, from 2006 to 2015, where he researched brain recovery after stroke, brain segmentation, and connectivity. Since 2021, he has been at Khalifa University, focusing on neuroimaging and bio-signal/image processing in diverse clinical populations. He has taught extensively, supervised several research projects, and serves as Editor-in-Chief of the International Journal of Imaging Systems and Technology and Handling Editor of the journal Imaging Neuroscience.